\documentclass[aps,pra,reprint,superscriptaddress,twocolumn,nofootinbib]{revtex4-2}
\usepackage{amsmath}
\usepackage{graphicx}
\usepackage[caption=false]{subfig}
\usepackage{fontawesome}
\usepackage{tikz}
  \usetikzlibrary{arrows.meta,calc,shapes,spy}
\usepackage{pifont}
  \newcommand{\vmark}{\text{\ding{51}}}
  \newcommand{\xmark}{\text{\ding{55}}}
\usepackage[hidelinks]{hyperref}

\begin{document}
\title{Symmetric observations without symmetric causal explanations}
\author{Christian William}
\affiliation{Department of Applied Physics, University of Geneva, 1211 Geneva 4, Switzerland}
\author{Patrick Remy}
\affiliation{Department of Applied Physics, University of Geneva, 1211 Geneva 4, Switzerland}
\author{Jean-Daniel Bancal}
\affiliation{Universit\'e Paris-Saclay, CEA, CNRS, Institut de physique th\'eorique, 91191, Gif-sur-Yvette, France}
\author{Yu Cai}
\affiliation{School of Physical and Mathematical Sciences, Nanyang Technological University, Singapore 637371}
\author{Nicolas Brunner}
\affiliation{Department of Applied Physics, University of Geneva, 1211 Geneva 4, Switzerland}
\author{Alejandro Pozas-Kerstjens}
\thanks{physics\faAt{}alexpozas.com}
\affiliation{Department of Applied Physics, University of Geneva, 1211 Geneva 4, Switzerland}

\begin{abstract}
    Inferring causal models from observed correlations is a challenging task, crucial to many areas of science.
    In order to alleviate the computational effort when sifting through possible causal explanations for some given observations, it is important to know whether symmetries in the observations correspond to symmetries in the underlying realization so that one can quickly discard impossible explanations.
    Via an explicit example, we demonstrate that, in general, symmetries cannot be exploited to reduce the hypothesis space.
    We use a tripartite probability distribution over binary events that is realized by using three (different) independent sources of classical randomness.
    We prove that even removing the condition that the sources distribute systems described by classical physics, the requirements that (i) the sources distribute the same physical systems, (ii) these physical systems respect relativistic causality, and (iii) the correlations are the observed ones are incompatible.
\end{abstract}

\maketitle

\section{Introduction}
From a fundamental point of view, understanding means uncovering the causal mechanisms that underlie the observed phenomena.
Despite its primary importance as a core component of the scientific process, an operational approach to causal inference has only been developed relatively recently \cite{causalitybook,spirtesbook}.

Modern approaches to causal inference, aided by improvements in computational capability, use discrete search techniques over the space of possible causal mechanisms \cite{zanga2022}.
A standard procedure is constraint based: Beginning with fully connected causal graphs, the independence relations extracted from the data are used to prune connections.
The paradigmatic example is the Peter-Clark algorithm \cite{spirtesbook}, which has since been refined and applied in multiple scenarios \cite{spirtes1995,colombo2014,le2019,pasquato2023}.
Alternatively, there also exist score-based techniques that rank each causal explanation by the likelihood that it can generate the observations \cite{akaike1974,geiger1994,jin2024}.
Both approaches suffer from \textit{curses of dimensionality}: For the former, the number of conditional independence tests to perform increases combinatorially with the number of variables; for the latter, one has to evaluate the score function in all possible causal graphs.
For any of the approaches it is thus very much desirable to restrict as much as possible the space of possible causal structures generating some given data.

\begin{figure}[t]
    \centering
    \subfloat[]{
        \scalebox{0.7}{
          \begin{tikzpicture}
            \draw [-{Latex[length=3mm]}] (2.6, 0) -- (4., 0);
            \draw [{Latex[length=3mm]}-] (0.5, 0) -- (2.1, 0);
            \draw [-{Latex[length=3mm]}] (1.125, 1.949) -- (0.25, 0.44);
            \draw [-{Latex[length=3mm]}] (1.125, 1.949) -- (1.95, 3.4);
            \draw [{Latex[length=3mm]}-] (2.55, 3.4) -- (3.375, 1.949);
            \draw [-{Latex[length=3mm]}] (3.375, 1.949) -- (4.25, 0.44);
            \node[draw,fill=white,circle,inner sep=5pt] at (2.25, 3.8) {\Large$A$};
            \node[draw,fill=white,circle,inner sep=5pt] at (0, 0) {\Large$B$};
            \node[draw,fill=white,circle,inner sep=5pt] at (4.5, 0) {\Large$C$};
            \node[draw,fill=white,regular polygon, regular polygon sides=3,inner sep=1.5pt] at (2.25, 0) {\Large$\alpha$};
            \node[draw,fill=white,regular polygon, regular polygon sides=3,inner sep=1.5pt] at (1.125, 1.949) {\Large$\gamma$};
            \node[draw,fill=white,regular polygon, regular polygon sides=3,inner sep=-0.2pt] at (3.375, 1.949) {\Large$\beta$};
          \end{tikzpicture}
      }
      \label{fig:triangleDAG}
    }
    \hspace{.2cm}
    \subfloat[]{
        \scalebox{0.7}{
          \begin{tikzpicture}
            % Top ovals
            \draw[fill=white] (0,2) ellipse (2cm and 1cm);
            \node at (-0.5,2.5) {\Large Models};
            \draw[fill=white] (0.6,1.7) ellipse (1cm and 0.5cm) node {\Large Symm.};
            
            % Bottom ovals
            \draw[fill=white] (0,-1) ellipse (2cm and 1cm);
            \node at (0,-1.5) {\Large Observations};
            \draw[fill=white] (-0.3,-0.7) ellipse (1cm and 0.5cm) node {\Large Symm.};
            
            % Curved arrows
            \draw[-{Latex[length=3mm]},>=stealth] (0.7,1.2) .. controls (0.7,0.5) .. (0.25,-0.3) node[midway,right] {\Large\vmark};
            \draw[-{Latex[length=3mm]},>=stealth] (-0.5,-0.2) .. controls (-0.4,0.5) .. (0.15,1.25) node[midway,left] {\Large\textbf{?}};
          \end{tikzpicture}
        }
        \label{fig:setting}
    }
    \caption{
        \textbf{Introduction to the work.} \protect\subref{fig:triangleDAG} The triangle causal structure.
        Three bipartite latent variables, $\alpha$, $\beta$ and $\gamma$, each influence a different set of two out of the three visible variables $A$, $B$ and $C$.
        This structure has received considerable attention in the context of nonlocality in quantum networks.
        \protect\subref{fig:setting} Pictorial representation of the relations between realizations and observations.
        In this work we demonstrate that $\textbf{?}=\xmark$.
        More specifically, there exist probability distributions over binary-valued observations that (i) are produced in the triangle causal structure by classical physical systems and (ii) are invariant under permutations of the visible nodes yet they do not admit explanations in terms of symmetric models, i.e., when $\alpha$, $\beta$ and $\gamma$ denote copies of the same physical system. 
        }
    \label{fig:intro}
\end{figure}

One of the most important ways of simplifying the analysis of phenomena is by understanding the operations or transformations that leave the phenomenon in question unaffected, i.e., by understanding its \textit{symmetries}.
In fact, symmetry has been the most important concept in the understanding of the fundamental laws of physics in the 20th century, and is the guide for further unification and progress \cite{gross1996}.
Symmetries allow to remove from the study all the redundancies due to, for instance, changes of reference frame or relabelings of the variables under consideration and focus on the essence of the phenomenon.
In the context of causal inference, symmetries can appear both at the level of observations (for example, when it is observed that the joint distribution of results of measurements on two systems remains the same under exchange of the systems) and at the level of the underlying causal model (i.e., when some parts of the causal mechanism can be exchanged without affecting the observations).
While the former can be extracted from the observed data, the latter are more subtle and uncovering them requires direct intervention on the process under study.

Therefore, in order to reduce as much as possible the search space in causal discovery algorithms, the question of understanding the level of symmetry in observations compared to the level of symmetry in the underlying model becomes central.
However, it is challenging to answer it, since observations and their realizations are very different objects.
Indeed, while it is intuitively clear (see Fig.~\ref{fig:setting} and the discussion below) that symmetries in a causal process lead to corresponding symmetries at the level of observable events, for some given observations with some symmetries there might be several causal explanations, and not all of them (if any) might have corresponding symmetries.

In this work we demonstrate that symmetries in observations cannot always be translated into corresponding symmetries in the causal mechanism that produces such observations.
As a consequence, causal discovery algorithms cannot in general use symmetries in the observed data to reduce the space of potential causal models for the data.
Since it is known that different physical theories allow us to produce different types of observations under the same causal mechanism \cite{bell1964}, we take a conservative approach and only demand consistency with relativistic causality.
Thus, our results are a proof that there exist observations that, despite presenting some observable symmetries, cannot be created by causal mechanisms that satisfy the corresponding symmetries in any physical theory consistent with relativistic causality.

\section{Observations versus realizations}
Our aim is to shed light on the relation between symmetries in observations and symmetries in the underlying causal model.
To do so, we focus on a simple scenario with three binary observed variables, where each pair is influenced by an independent latent variable (see Fig.~\ref{fig:triangleDAG}).
This is known as the triangle scenario and has received considerable attention in the context of Bell nonlocality \cite{TavakoliPozas2022,fritz2012,renou2019,fraser2018,gisin2020,pozas2023}.

In the context of Bayesian probability theory, all nodes in Fig.~\ref{fig:triangleDAG} represent random variables, and the arrows indicate a functional dependence.
Under the lens of quantum information theory, this is precisely the physical realization of network local models \cite{fritz2012,TavakoliPozas2022}, i.e., this interpretation of the nodes and arrows leads to the characterization of the observations that are generated by measuring physical systems whose dynamics are described by classical mechanics \cite{wood2015}.
This presented an opportunity for an extended interpretation of the graphs such as that in Fig.~\ref{fig:triangleDAG}, whereby the latent nodes describe physical systems that are distributed to parties, represented by the visible nodes \cite{Henson2014,Weilenmann2020}.
On the received physical systems, each party performs a measurement in order to produce the concrete value of their variable in a given round.

The concrete expression for the calculation of the probabilities of the different events, thus, has a dependence on the physical theory assumed to govern the systems distributed.
For classical systems, the expression for calculating probabilities of observable events is given by the Markov condition.\footnote{The Markov condition states that the value of any variable in the graph depends only on the value of the variables that have a direct influence on it.}
In the case of the systems being described by quantum mechanics, it is Born's rule.
These are nothing but particular cases of what has evolved into an information-theoretic perspective of physics, known as generalized probabilistic theories \cite{janotta2014,plavala2023,Weilenmann2020}.
In this theory-agnostic formulation, observations created by the causal mechanism in Fig.~\ref{fig:triangleDAG} are described by
\begin{equation}
    \begin{aligned}
        p(A&=a,B=b,C=c) \\
        &= e^a_{A_1A_2}\otimes f^b_{B_1B_2}\otimes g^c_{C_1C_2}\left(\alpha_{B_2C_1}\otimes\beta_{C_2A_1}\otimes\gamma_{A_2B_1}\right),
        \end{aligned}
    \label{eq:GPT}
\end{equation}
where the subscripts denote vector spaces; $\alpha$, $\beta$ and $\gamma$ (the \textit{states} distributed by the latent nodes) are elements of a tensor product of the corresponding spaces, and $e^a$, $f^g$ and $g^c$ (the \textit{effects} that describe the measurements performed by the parties) are elements of the dual of the tensor product of the corresponding spaces, all satisfying suitable properties \cite{ludwigbook,krausbook}.
From now on, we will use the shorthand notation $p(a,b,c)=p(A=a,B=b,C=c)$.

Consider now the symmetric version of the scenario, where the three latent nodes distribute copies of the same state, $\omega$, and all the parties perform the same effect, $e$ (see Fig.~\ref{fig:triangle}).
The model for distributions observed in this situation has the form
\begin{equation}
    p(a,b,c)\,{=}\,e^a_{A_1A_2}\!\otimes e^b_{B_1B_2}\!\otimes e^c_{C_1C_2} \!\left(\omega_{B_2C_1}\,{\otimes}\,\omega_{C_2A_1}\,{\otimes}\,\omega_{A_2B_1}\right)\!.
    \label{eq:symmGPT}
\end{equation}
Importantly, in this equation, permuting the labels $A$, $B$ and $C$ leads to the exact same expression.

Now consider the parametrization of the distributions above.
If the visible variables are binary, $a,\,b,\,c\in\{-1,1\}$, an arbitrary probability distribution $p(a,b,c)$ is characterized by seven independent parameters, namely, all probabilities minus the normalization constraint.
However, for distributions that are generated via Eq.~\eqref{eq:symmGPT}, the symmetry reduces this number to just three parameters: Every tripartite distribution over binary variables that is symmetric under permutation of the parties can be expressed as
	\begin{equation}
		p(a,b,c)=\frac18 [1+(a+b+c)E_1+(ab+bc+ca)E_2+abcE_3]
		\label{eq:prob}
	\end{equation}
for some $E_1$, $E_2$, $E_3\in[-1,1]$.
These variables correspond to the single-, two- and three-body correlators of the distribution, respectively.
In short, the causal model given by Eq.~\eqref{eq:symmGPT} implies that the observed distribution can be expressed via Eq.~\eqref{eq:prob}.
This is pictorially captured in the downward-pointing arrow in Fig.~\ref{fig:setting}.

Here we ask the opposite question: If we have a realization (in the sense of Eq.~\eqref{eq:GPT}) that produces a distribution invariant under permutation of parties (i.e., of the form of Eq.~\eqref{eq:prob}), can we always find an alternative symmetric realization of the form of Eq.~\eqref{eq:symmGPT}?
Previous works have touched upon this question in the context of classical realizations \cite{gisin2020,silva2023}.
Surprisingly, answers point towards opposite conclusions: On the one hand, the classical model (given in the Supplementary Note 2 of \cite{gisin2020}) for the probabilistic mixture of a uniform shared random bit with the least amount of white noise that makes the distribution known to be producible in the triangle scenario with classical sources is symmetric.
On the other hand, the analogous study for the distribution where uniformly all but one of the parties produce a given output (known as the $W$ distribution) showed that in this case the model was asymmetric \cite{silva2023}, reporting not being able to achieve the same results even with more general symmetric models.

Our work settles this apparent discrepancy by proving a strong version of the latter observation.
More specifically, we give a family of tripartite probability distributions over binary variables that (i) can be produced in the triangle scenario when the sources distribute systems described by classical physics and (ii) are invariant under permutations of parties, which we prove are impossible to reproduce with symmetric strategies and any physical systems that are consistent with relativity.
This includes theoretical stronger-than-quantum systems \cite{Henson2014,janotta2014,plavala2023}.

\section{Proof via counterexample} 
Below we demonstrate that symmetries in the observed statistics cannot be exploited to assume analogous symmetries in the causal models that produce those statistics.
In order to do so, a single counterexample suffices.
The example that we use comes from the study of quantum nonlocality in networks \cite{TavakoliPozas2022} in the triangle scenario.
It is the family of distributions of the form given in Eq.~\eqref{eq:prob} with $E_1=E_1^c\approx0.1753$ (the exact form of $E_1^c$, which is the largest value of $E_1$ for which a distribution with $E_2=-1/3$ is known to have a triangle-local model, is given in Appendix \ref{app:E1c}) and $E_2=-1/3$.
For $E_2=-1/3$, the positivity of Eq.~\eqref{eq:prob} for any $a,b$ and $c$ fixes $E_3=-3E_1$, so in the case that we will consider we will have $E_3\approx-0.5260$.
However, our proof technique does not use the information about $E_3$, so when we consider different values of $E_1$ and $E_2$ later on, the proofs extracted will hold for any value of $E_3$ that leads to a well-defined probability distribution.

We will prove the impossibility of generating the distribution \eqref{eq:prob} for $E_1=E_1^c$ and $E_2=-1/3$ with symmetric realizations in the triangle scenario via reduction to the absurd.
Thus, let us initially assume that the distribution admits a realization, i.e., that there exists a source of bipartite physical systems $\omega$ and a measurement device characterized by the effect $e$ that produces a classical bit out of processing two input systems, which, when three copies of each are arranged in the triangle-like structure of Fig.~\ref{fig:triangle}, the possible outcomes are produced according to Eq.~\eqref{eq:prob} with $E_1=E_1^c$ and $E_2=-1/3$.
If that were indeed the case, one could take not just three but an arbitrary number of copies of the source and the measurement device and arrange them to form a polygon (see Fig.~\ref{fig:inflation}).
In the following we choose to use seven copies of the sources and of the measurement devices, creating a heptagon-like structure.
We choose this number because, for this specific distribution, it is the smallest number of copies that leads to a contradiction. 
For other distributions or structures, the smallest number of copies needed to observe a contradiction may vary.

\begin{figure}[t]
    \centering
    \subfloat[]{
        \begin{tikzpicture}
            \node[] at (0,0) {};
            \draw[thick,rotate=90,xshift=-1.1cm] (0:1) \foreach \x in {120,240} { -- (\x:1) } -- cycle;
            \foreach \x in {0,120,240} {
                \fill[black,rotate=90,xshift=-1.1cm] (\x:1) circle (2pt);
            }
        \end{tikzpicture}
        \label{fig:triangle}
    }
    \hspace{1cm}
    \subfloat[]{
        \begin{tikzpicture}
            \draw[thick,rotate=40.5] (0:1) \foreach \x in {51,102,...,309} { -- (\x:1) } -- cycle;
            \foreach \x in {0,51,102,...,309} {
                \fill[black,rotate=40.5] (\x:1) circle (2pt);
            }
        \end{tikzpicture}
        \label{fig:inflation}
    }
    \caption{
        \textbf{Inflation for realizations invariant under permutations of parties.} \protect\subref{fig:triangle} Simplified representation of the triangle causal structure in Fig.~\ref{fig:triangleDAG} for symmetric causal scenarios.
        Each line represents an independent copy of the same bipartite state $\omega$ that is distributed to the two parties (the nodes) that it is connected to.
        All parties perform the same measurement, given by the effect $\{e^o\}_o$, to produce their outcomes.
        Any process following this causal mechanism leads to distributions over the observable events that are invariant under permutations of the observable nodes.
        For all observations that are produced in this way, one can consider the distributions that are created when using more copies of the state and the measurement device like, for instance, that in \protect\subref{fig:inflation}.
        We use the consistency conditions implied by the existence of this distribution to prove that the distribution with $E_1=E_1^c$, $E_2=-1/3$, and $E_3=E_3^c$ is impossible to generate in the symmetric triangle network under any physical theory consistent with relativity.
        }
\end{figure}

In this new heptagon-like structure, the measurements produce outcomes $a_0,\dots,a_6\in\{-1,1\}^7$ according to a distribution $p_\text{inf}(a_0,\dots,a_6)$.
This distribution has several properties.
First, since it is a probability distribution, all the probabilities must be non-negative and sum to 1.
Second, since all the sources and measurement devices are the same, it is invariant under any cyclic permutation of the parties, i.e.,
\begin{equation}
    p_\text{inf}(a_0,\dots,a_6)=p_\text{inf}(a_{i},\dots,a_{6\oplus i})\quad\forall i=1,\dots,6,
    \label{eq:symm}
\end{equation}
where the $\oplus$ symbol represents addition modulo 7.
Third, since the sources and measurement devices used in creating $p_\text{inf}$ are the same as those used to create $p$, the two-body marginals (and hence the one-body marginals) of both distributions must be the same.

This implies a number of relations.
Take, for instance, the marginalization over sites $i$ and $i+2$.\footnote{Due to the symmetry constraints \eqref{eq:symm}, the concrete value of $i$ is not important.}
In Fig.~\ref{fig:inflation} one sees that this five-partite marginal has two independent components: one corresponding to a single-body marginal, and another one corresponding to a four-body marginal.
Since the sources and the measurement devices used in Fig.~\ref{fig:inflation} are the same as those used in Fig.~\ref{fig:triangle} (i.e., the state $\omega$ and the effect $e$ in Eq.~\eqref{eq:symmGPT}), the single-body marginal has to be the same as that given by marginalizing Eq.~\eqref{eq:prob}.
This leads to the relation
\begin{equation}
    \begin{aligned}
        \sum_{a_0,a_2}&p_\text{inf}(a_0,a_1,a_2,a_3,a_4,a_5,a_6)\\
        &=p_{\text{inf},1}(a_1)p_{\text{inf},4}(a_3,a_4,a_5,a_6)\\
        &=\frac{1}{2}\left(1+a_1 E_1\right)\sum_{a'_0,a'_1,a'_2}p_\text{inf}(a'_0,a'_1,a_2',a_3,a_4,a_5,a_6)\\
        &\hspace{4.5cm}\forall a_1,a_3,a_4,a_5,a_6.
    \end{aligned}
    \label{eq:iden1}
\end{equation}

Similarly, one can consider the five-partite distribution that results from marginalizing sites $i$ and $i+3$.
In this case, the marginal has also two independent components: one corresponding to a bipartite marginal (which, due to the construction, has to coincide with the two-body marginalization of Eq.~\eqref{eq:prob}) and the other one corresponding to a three-party marginal.
The relation obtained is thus
\begin{equation}
    \begin{aligned}
        \sum_{a_0,a_3}&p_\text{inf}(a_0,a_1,a_2,a_3,a_4,a_5,a_6)\\
        &=p_{\text{inf},2}(a_1,a_2)p_{\text{inf},3}(a_4,a_5,a_6)\\
        &=\frac{1}{4}\left[1+\left(a_1+a_2\right)E_1+a_1 a_2E_2\right]\\
        &\qquad\times\!\!\!\!\sum_{a'_0,a'_1,a'_2,a'_3}\!\!\!\!p_\text{inf}(a'_0,a_1',a_2',a_3',a_4,a_5,a_6)\\
        &\hspace{4.5cm}\forall a_1,a_2,a_4,a_5,a_6.
    \end{aligned}
    \label{eq:iden2}
\end{equation}
These two relations are sufficient to derive all other ones that can be obtained in the network.

For a fixed target distribution, i.e., for fixed $E_1$ and $E_2$, Eqs.~\eqref{eq:symm}-\eqref{eq:iden2} are linear in the probabilities $p_\text{inf}(a_0,\dots,a_6)$.
This means that determining the existence of a set of numbers $\{p_\text{inf}(a_0,\dots,a_6)\}$ that satisfies the necessary properties is an instance of a linear program \cite{boydbook}.
In the computational appendix \cite{compapp} we set up this linear program for $E_1=E_1^c$ and $E_2=-1/3$ and show that it is infeasible, i.e., that in every set of numbers $\{p_\text{inf}(a_0,\dots,a_6)\}_{a_0,\dots,a_6\in\{-1,1\}}$ that simultaneously sum to one and satisfy Eqs.~\eqref{eq:symm}-\eqref{eq:iden2} at least one of the numbers is negative, and thus they cannot form a probability distribution.
An alternative way to interpret this is that the constraints of positivity, normalization, symmetry, and matching marginals constrain the set of allowed values for $E_1$ and $E_2$.
In Sec.~\ref{sec:cert} and Appendix~\ref{app:cert} we take this perspective to derive Bell-like inequalities whose violation certifies the impossibility of a symmetric realization.
More sepcifically, we take positive linear combinations of the $p_\text{inf}(a_0,\dots,a_6)$ (which thus should always evaluate to a non-negative quantity if all the $p_\text{inf}(a_0,\dots,a_6)$ are non-negative) and show that for $E_1=E_1^c$ and $E_2=-1/3$ the combination evaluates to a negative quantity.

Recall that at least one collection of non-negative $\{p_\text{inf}(a_0,\dots,a_6)\}_{a_0,\dots,a_6\in\{-1,1\}}$ that sum to 1 and satisfy Eqs.~\eqref{eq:symm}-\eqref{eq:iden2} should exist if the original distribution $p(a,b,c)$ given in Eq.~\eqref{eq:prob} was possible to create with a symmetric strategy in the triangle scenario.
Thus, we have a proof that the probability distribution given by Eq.~\eqref{eq:prob} for $E_1=E_1^c$ and $E_2=-1/3$ does not admit a symmetric realization.

At this point, two facts must be highlighted: (i) It is known that the distribution \eqref{eq:prob} with $E_1=E_1^c$, $E_2=-1/3$ and $E_3=-3E_1^c$ can be realized with a non-symmetric strategy when the sources distribute physical systems obeying the rules of classical physics (see Appendix \ref{app:E1c} and Ref.~\cite{gisin2020}) and (ii) we have not required any special property of the sources in our symmetric construction.
In particular, we have not demanded that they distribute physical systems described according to any particular physical theory.
Thus, our result is a proof that symmetries in a probability distribution do not always translate to symmetries in its realization: While Eq.~\eqref{eq:prob} with $E_1=E_1^c$, $E_2=-1/3$ and $E_3=-3E_1^c$ is symmetric under permutations of parties and can be realized in the triangle scenario classically, neither by using physical systems described by classical, quantum, or potentially stronger-than-quantum physics, it will be possible to realize in setups where all sources and measurement devices are equal.

\section{Systematic approach}
The proof above is an instance of the inflation argument.
Inflation has turned out to be a powerful tool for characterizing the probability distributions that can be produced in causal scenarios with latent nodes that distribute classical \cite{wolfe2019,pozas2022b,pozas2023,fraser2018,lauand2024,lauand2023}, quantum \cite{wolfe2021,pozas2023}, and arbitrary physical systems \cite{gisin2020,camillo2023}, as well as scenarios with different sources being of different types \cite{pozas2022,wang2023}.
Its success is due, in part, to its applicability to any causal structure, and in part due to the fact that it relaxes the sets of correlations under study to forms amenable to linear or semidefinite programming \cite{TavakoliPozas2024} that can be solved in many cases with standard computational resources \cite{boghiu2022,TavakoliPozas2024}.
Moreover, it is possible to define hierarchies of inflation relaxations \cite{navascues2017,wolfe2021,ligthart2023}, which in some cases are known to converge \cite{navascues2017,ligthart2023,ligthart2023b}.

In our case of interest, it is possible to define an inflation hierarchy of compatibility conditions with symmetric realizations, which we pictorially illustrate in Fig.~\ref{fig:hierarchy}.
Step $n$ in the hierarchy demands the existence of $n$ probability distributions $\{p_k(a_0,\dots,a_{k-1})\}_{k=4}^{n+3}$, all of them symmetric under cyclic permutations, i.e., satisfying the analogs of Eq.~\eqref{eq:symm}, and each of them compatible (in the sense of Eqs.~\eqref{eq:iden1} and \eqref{eq:iden2}) with the distribution under test \cite{girardin2023}.
Since all the sources and measurement devices in all inflations are required to be identical, the distributions are related to each other via
\begin{equation}
    \sum_{a_{k-2},a_{k-1}}p_k(a_0,\dots,a_{k-1}) = \sum_{a_{k-2}}p_{k-1}(a_0,\dots,a_{k-2})
\end{equation}
for all $a_0,\dots,a_{k-3}$.

\begin{figure}[t]
    \centering
    \begin{tikzpicture}
    
    % Colors
    \definecolor{lightgray}{gray}{0.8}
    \definecolor{mediumgray}{gray}{0.7}
    \definecolor{darkgray}{gray}{0.6}
    
    % Rectangles
    \fill[darkgray,rounded corners] (-0.45,-0.45) rectangle (7.2,2.75);
    \fill[mediumgray,rounded corners] (-0.35,-0.35) rectangle (4.6,2.5);
    \fill[lightgray,rounded corners] (-0.25,-0.25) rectangle (2,2.25);
    \draw[thick] (0,0) -- (1.5,0) -- (1.5,1.5) -- (0,1.5) -- cycle;

    % Square
    \foreach \x in {0,1.5} {
        \foreach \y in {0,1.5} {
            \fill[black] (\x,\y) circle (2pt);
        }
    }
    
    % Pentagon
    \draw[thick,xshift=3.2cm,yshift=0.8cm,rotate=18] (0:1) \foreach \x in {72,144,...,288} { -- (\x:1) } -- cycle;
    \foreach \x in {0,72,...,288} {
        \fill[black,xshift=3.2cm,yshift=0.8cm,rotate=18] (\x:1) circle (2pt);
    }
    
    % Hexagon
    \draw[thick,xshift=5.9cm,yshift=0.85cm] (0:1) \foreach \x in {60,120,...,300} { -- (\x:1) } -- cycle;
    \foreach \x in {0,60,...,300} {
        \fill[black,xshift=5.9cm,yshift=0.85cm] (\x:1) circle (2pt);
    }

    % Labels
    \node[] at (7.7,0.8) {\LARGE$\dots$};
    \node[] at (0.8,2) {\large$n=1$};
    \node[] at (3.2,2.2) {\large$n=2$};
    \node[] at (5.9,2.2) {\large$n=3$};
    
    \end{tikzpicture}
    \caption{
        \textbf{Hierarchy of inflations used.}
        In each level, we require the existence of probability distributions for each of the networks with a smaller number of nodes as well.
        These distributions are related to each other by constraints on the marginals over the same number of nodes (see Eqs.~\eqref{eq:iden1} and \eqref{eq:iden2}).}
    \label{fig:hierarchy}
\end{figure}

Using a standard tabletop computer (approximately 1 GB of RAM) it is possible to run up to the 11th level of the hierarchy (i.e., up to the inflation with 14 nodes) in a few seconds per distribution, producing the orange area in Fig.~\ref{fig:results}, which denotes symmetric distributions that do not admit symmetric realizations.
Using approximately 50 GB of RAM and around one day of computation time, we are able to certify that all distributions of the form of Eq.~\eqref{eq:prob} for $E_2=-1/3$ are impossible to realize with arbitrary symmetric strategies for $E_1\geq0.1580$ (the red circle in Fig.~\ref{fig:results}), using the 15th level of the hierarchy.
The code executed can be found in the computational appendix \cite{compapp}.

\section{Certificates of incompatibility}
\label{sec:cert}
If one wants to assess the (in)compatibility of multiple distributions, a possibility is to run the corresponding linear programs for each of them.
However, this way of proceeding is inherently not robust to noise, since the linear program is specifically tailored to a particular distribution via Eqs.~\eqref{eq:iden1} and \eqref{eq:iden2}.
For practical application, it is more useful to have certificates (sometimes also known as witnesses) of incompatibility, which are in general smooth functions of the distribution that are positive for the whole set of compatible distributions.
As a consequence, its evaluation to a negative value constitutes a proof of incompatibility.
Certificates of incompatibility are commonplace in quantum information theory, where they are used to certify, e.g., entanglement \cite{Terhal2000,Lewenstein2001,Chruscinski2014}, nonlocality \cite{Brunner2014,Designolle2024}, or causal structure \cite{wolfe2021,Ulibarrena2024}.
In the context of inflation, these certificates of infeasibility can generally be understood as Bell-like inequalities \cite{bell1964} that separate (relaxations of) the set of distributions that can be produced in a given scenario from (a subset of) those that cannot.

\begin{figure}[ht]
    \centering
    \begin{tikzpicture}[spy using outlines={rectangle, magnification=1.85, width=4cm, height=2cm, connect spies}]
        \node[anchor=south west, inner sep=0] (image) at (0,0) {\includegraphics[width=0.9\columnwidth]{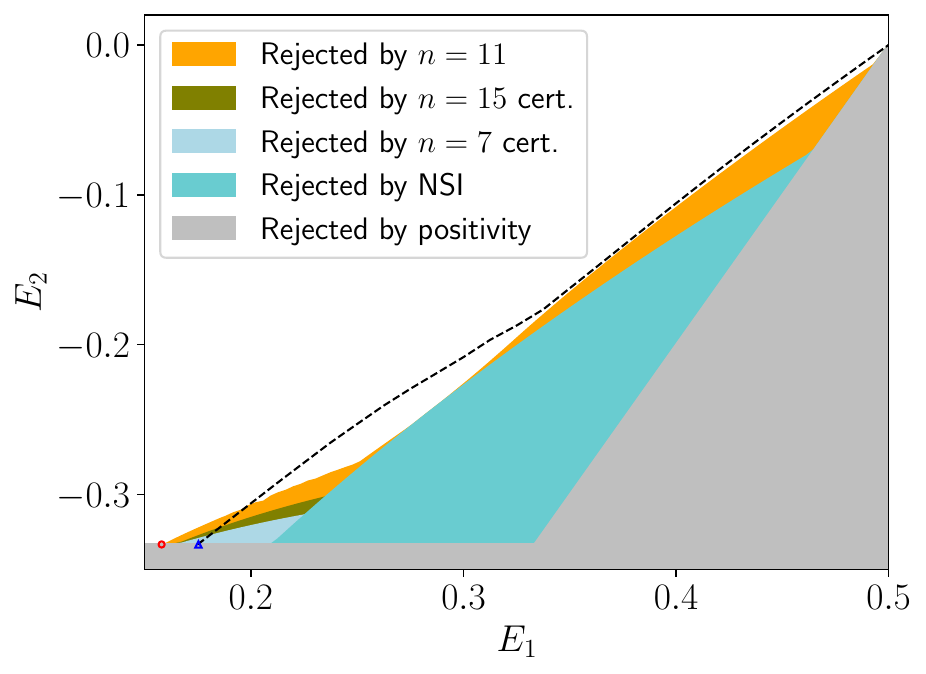}};
        \spy[black] on (2.2, 1.35) in node [fill=white] at (6.5, 1.95);
    \end{tikzpicture}
    \caption{\textbf{Symmetric distributions without symmetric realizations.}
    Projection of the space of probability distributions given by Eq.~\eqref{eq:prob} on the plane defined by $E_1$ and $E_2$.
    The points in the gray area produce negative probabilities for any value of $E_3$.
    The points in the blue area correspond to distributions that cannot be generated in the triangle scenario, regardless of the nature of the sources \cite{gisin2020}.
    In the area to the left of the black dashed line, for every $E_1$ and $E_2$ there exists at least one value of $E_3$ for which the corresponding distribution admits a realization in the triangle structure with classical latent variables \cite{gisin2020}.
    In particular, the blue triangle denotes the distribution that we use in the work, with $E_1=E_1^c\approx0.1753$ and $E_2=-1/3$.
    The light blue area depicts the distributions that violate the compatibility witness of Eq.~\eqref{eq:7cert}.
    The orange area denotes the distributions that are proven not to admit a symmetric realization in the triangle scenario using the 11th level of the corresponding inflation hierarchy using the constraints of the type of Eqs.~\eqref{eq:iden1} and \eqref{eq:iden2}.
    Using approximately 50 GB of RAM and one day of computation time, we are able to identify as incompatible also the point shown with a red circle ($E_1=0.1580$ and $E_2=-1/3$) using the 15th level of the hierarchy.
    The dark green area contains the distributions that are identified as incompatible by the certificate obtained by replacing the constraints \eqref{eq:iden1} and \eqref{eq:iden2} by \eqref{eq:nolpi} and testing the distribution with $E_1=0.1656$ and $E_2=-1/3$.
    The expression of the certificate is given in Appendix \ref{app:cert}.
    }
    \label{fig:results}
\end{figure}

A possible way to obtain certificates of incompatibility is to perform variable elimination on the set of positivity constraints $\{p_\text{inf}(a_0,\dots,a_k)\geq 0\}_{a_0,\dots,a_k}$ using only positive coefficients, since by definition these combinations are positive for all probability distributions compatible with the inflation under consideration.
By doing so in the seven-sided inflation described in Fig.~\ref{fig:inflation} (see Appendix \ref{app:elimination} for details), one arrives at the fact that all distributions that admit a symmetric realization and for which $6E_1+5E_2+1\geq 0$, $13E_1-1\geq 0$ and $13E_1^2-10E_2-3\geq0$ are true satisfy
\begin{equation}
    \begin{aligned}
        28 E_1^4 + 7 E_1^2 E_2^2 + 42 E_1^2 E_2 + 7 E_1^2 + 28 E_1 E_2^2 & \\
        + 56 E_1 E_2 + 12 E_1 + 7 E_2^3 + 3 E_2 + 2&\geq 0.
    \end{aligned}
    \label{eq:7cert}
\end{equation}
The auxiliary conditions $6E_1+5E_2+1\geq 0$, $13E_1-1\geq 0$ and $13E_1^2-10E_2-3\geq0$ arise from the elimination process and limit the space of distributions that can be proved incompatible.
However, as we show in Fig. \ref{fig:7cert} in Appendix \ref{app:elimination}, for the case under study they do not restrict significantly the set of distributions that can be tested.
When substituting $E_1=E_1^c\approx0.1753$ and $E_2=-1/3$ we obtain that $6E_1+5E_2+1\approx0.3851$, $13E_1-1\approx1.2789$, and $13E_1^2-10E_2-3\approx0.7328$, while Eq.~\eqref{eq:7cert} evaluates to approximately $-0.048$.
We depict the set of distributions that violate Eq.~\eqref{eq:7cert} in light blue in Fig.~\ref{fig:results}.

The process of variable elimination becomes tedious as the number of elements in the probability distribution increases.
In order to find certificates of incompatibility in a more systematic manner, we can use 
Farkas' lemma, which states that, for every linear program, either a solution exists or there exists a certificate (findable by linear programming) that there is no solution.

The constraints \eqref{eq:iden1} and \eqref{eq:iden2} are known in the literature as linearized polynomial constraints \cite{pozas2022b} and they are examples of constraints that lead to certificates of infeasibility that are only applicable to the distribution under scrutiny \cite{pozas2022b}.
In order to obtain certificates that apply to other distributions, one can either follow the prescriptions of Ref.~\cite{pozas2022b} or alternatively relax Eqs.~\eqref{eq:iden1} and \eqref{eq:iden2} to constraints on only the marginals that can be associated with (products of) marginals in the original network.
For instance, in the inflation in Fig.~\ref{fig:inflation}, these constraints are
\begin{equation}
    \begin{aligned}
        \sum_{a_2,a_5,a_6}\!\! p_\text{inf}(a_0,\dots,a_6)=\frac{1}{16} &\left[1+(a_0+a_1)E_1+a_0a_1E_2\right]\\
        \times&\left[1+(a_3+a_4)E_1+a_3a_4E_2\right],\\
        \sum_{a_2,a_4,a_6}\!\! p_\text{inf}(a_0,\dots,a_6) =\frac{1}{16} &\left[1+(a_0+a_1)E_1+a_0a_1E_2\right]\\
        \times&\left(1+a_3E_1\right)\left(1+a_5E_1\right).
    \end{aligned}
    \label{eq:nolpi}
\end{equation}
Replacing the constraints \eqref{eq:iden1} and \eqref{eq:iden2} with \eqref{eq:nolpi} and running the respective linear programs, we are able to determine the infeasibility of distributions for $E_2=-1/3$ only for $E_1\geq0.1656$ using the 15th level of the hierarchy.
In exchange, we obtain a witness of incompatibility with symmetric realizations in the triangle, which is the upper boundary to the dark green region in Fig.~\ref{fig:results}.
The expression of this certificate is given in Appendix \ref{app:cert} and stored in machine-readable form in the computational appendix \cite{compapp}.
Despite the higher level in the hierarchy, the set of distributions identified as incompatible with this certificate is smaller than the set identified by the constraints \eqref{eq:iden1} and \eqref{eq:iden2} (in orange).
This confirms that the constraints \eqref{eq:nolpi} are notably weaker than \eqref{eq:iden1} and \eqref{eq:iden2}, a fact that is now well known in the literature \cite{pozas2023,pozas2022b,AlexThesis,plavala2025}.

\section{Discussion}
In physics, symmetries allow us to gain a better understanding of the concrete process under study.
However, there exist cases where the symmetries appear to be emergent, in the sense that the macroscopic observations have a symmetry that is not present in the currently available microscopic model of the phenomenon.
An example of this can be found in the context of hydrodynamic turbulence, where some phenomena appear to have scale invariance despite the underlying Navier-Stokes equations not possessing such symmetry \cite{frisch1995turbulence,shavit2022}.

The results in this work demonstrate a strong version of these observations: There exist distributions over measurable events that have a macroscopic symmetry (in our case, under permutations of the measured variables), for which we prove that a realization in terms of a symmetric causal model is impossible in any physical theory consistent with relativistic causality.
In other words, the conditions of symmetry in observable events, symmetry in the causal mechanism that produces such events, and relativistic causality are not always all compatible.

From a pure causal inference perspective, when looking for causal explanations of observations, it is desirable to reduce the space of potential causal explanations as much as possible.
Our results indicate that symmetries in the observed data cannot be used for this purpose, at least in the general case.
An important question that our work motivates is whether this is always the case or if, on the contrary, there exist specific symmetries, scenarios, or combinations of them in which the equivalence between observations and their realizations holds.
On a more positive side, we have built a hierarchy of simple tests that may be useful for certification purposes.
The reason for this is that the failure of any of them constitutes a proof that the observed statistics cannot be produced in the model under scrutiny.
This may have applications in cryptography, where the parties may want to guarantee some properties of the setup while potentially not trusting each other.

While illustrated in this work for the triangle causal structure, an analogous approach can be followed for more complex scenarios.
One could also follow the same procedure to impose symmetries only in specific parts of the model, in analogy with the approach that is followed to characterize the distributions generated in scenarios where the different sources distribute systems described by different physical theories \cite{wolfe2021,pozas2022}.
Following this line of research, it would be interesting to characterize more fine-grained, physically relevant symmetries.
Two such symmetries are sources that distribute states that are invariant under permutations of the subsystems (in quantum mechanics, examples of these states are the Bell states, or the Greenberger-Horne-Zeilinger and $W$ states for multipartite systems) and measurement devices that do not distinguish between input ports.
In inflations of loop scenarios (such as the triangle network), the first symmetry is reflected as invariance under mirror symmetry with respect to the symmetric source and the second one as invariance under mirror symmetry with respect to the party that holds the symmetric measurement device.
For illustration, take a six-partite inflation of the triangle network, where now we do not assume models that are symmetric under permutations of parties.
If, say, the source between parties $A$ and $B$ distributes symmetric states, then the inflation satisfies $p_\text{inf}(a_1,b_1,c_1,a_2,b_2,c_2)=p_\text{inf}(b_1,a_1,c_2,b_2,a_2,c_1)$, and if the measurement of device $B$ does not distinguish between input ports the inflation satisfies $p_\text{inf}(a_1,b_1,c_1,a_2,b_2,c_2)=p_\text{inf}(c_1,b_1,a_1,c_2,b_2,a_2)$.
How these symmetries are imposed in more complex causal structures, and whether more symmetries can be imposed in the models, is left to future work.

\acknowledgments
We thank Fernando Parisio and Bastien Chopard for helpful discussions and comments.
This work was supported by the National Research Foundation, Singapore under its Quantum Engineering Programme (National Quantum-Safe Network NRF2021-QEP2-04-P01) and the Swiss National Science Foundation (grant numbers 192244 and 224561).
Computations were performed in part at the University of Geneva using the Baobab HPC service.

\bibliography{references}

\begin{thebibliography}{52}%
\makeatletter
\providecommand \@ifxundefined [1]{%
 \@ifx{#1\undefined}
}%
\providecommand \@ifnum [1]{%
 \ifnum #1\expandafter \@firstoftwo
 \else \expandafter \@secondoftwo
 \fi
}%
\providecommand \@ifx [1]{%
 \ifx #1\expandafter \@firstoftwo
 \else \expandafter \@secondoftwo
 \fi
}%
\providecommand \natexlab [1]{#1}%
\providecommand \enquote  [1]{``#1''}%
\providecommand \bibnamefont  [1]{#1}%
\providecommand \bibfnamefont [1]{#1}%
\providecommand \citenamefont [1]{#1}%
\providecommand \href@noop [0]{\@secondoftwo}%
\providecommand \href [0]{\begingroup \@sanitize@url \@href}%
\providecommand \@href[1]{\@@startlink{#1}\@@href}%
\providecommand \@@href[1]{\endgroup#1\@@endlink}%
\providecommand \@sanitize@url [0]{\catcode `\\12\catcode `\$12\catcode
  `\&12\catcode `\#12\catcode `\^12\catcode `\_12\catcode `\%12\relax}%
\providecommand \@@startlink[1]{}%
\providecommand \@@endlink[0]{}%
\providecommand \url  [0]{\begingroup\@sanitize@url \@url }%
\providecommand \@url [1]{\endgroup\@href {#1}{\urlprefix }}%
\providecommand \urlprefix  [0]{URL }%
\providecommand \Eprint [0]{\href }%
\providecommand \doibase [0]{https://doi.org/}%
\providecommand \selectlanguage [0]{\@gobble}%
\providecommand \bibinfo  [0]{\@secondoftwo}%
\providecommand \bibfield  [0]{\@secondoftwo}%
\providecommand \translation [1]{[#1]}%
\providecommand \BibitemOpen [0]{}%
\providecommand \bibitemStop [0]{}%
\providecommand \bibitemNoStop [0]{.\EOS\space}%
\providecommand \EOS [0]{\spacefactor3000\relax}%
\providecommand \BibitemShut  [1]{\csname bibitem#1\endcsname}%
\let\auto@bib@innerbib\@empty
%</preamble>
\bibitem [{\citenamefont {Pearl}(2009)}]{causalitybook}%
  \BibitemOpen
  \bibfield  {author} {\bibinfo {author} {\bibfnamefont {J.}~\bibnamefont
  {Pearl}},\ }\href {https://doi.org/10.1017/CBO9780511803161} {\emph {\bibinfo
  {title} {{Causality: Models, Reasoning, and Inference}}}}\ (\bibinfo
  {publisher} {Cambridge University Press},\ \bibinfo {year}
  {2009})\BibitemShut {NoStop}%
\bibitem [{\citenamefont {Spirtes}\ \emph {et~al.}(2011)\citenamefont
  {Spirtes}, \citenamefont {Glymour},\ and\ \citenamefont
  {Scheines}}]{spirtesbook}%
  \BibitemOpen
  \bibfield  {author} {\bibinfo {author} {\bibfnamefont {P.}~\bibnamefont
  {Spirtes}}, \bibinfo {author} {\bibfnamefont {C.}~\bibnamefont {Glymour}},\
  and\ \bibinfo {author} {\bibfnamefont {R.}~\bibnamefont {Scheines}},\ }\href
  {https://doi.org/10.1007/978-1-4612-2748-9} {\emph {\bibinfo {title}
  {Causation, Prediction, and Search}}},\ Lecture Notes in Statistics\
  (\bibinfo  {publisher} {Springer New York},\ \bibinfo {year}
  {2011})\BibitemShut {NoStop}%
\bibitem [{\citenamefont {Zanga}\ \emph {et~al.}(2022)\citenamefont {Zanga},
  \citenamefont {Ozkirimli},\ and\ \citenamefont {Stella}}]{zanga2022}%
  \BibitemOpen
  \bibfield  {author} {\bibinfo {author} {\bibfnamefont {A.}~\bibnamefont
  {Zanga}}, \bibinfo {author} {\bibfnamefont {E.}~\bibnamefont {Ozkirimli}},\
  and\ \bibinfo {author} {\bibfnamefont {F.}~\bibnamefont {Stella}},\
  }\bibfield  {title} {\bibinfo {title} {A survey on causal discovery: Theory
  and practice},\ }\href
  {https://doi.org/https://doi.org/10.1016/j.ijar.2022.09.004} {\bibfield
  {journal} {\bibinfo  {journal} {Int. J. Approx. Reason.}\ }\textbf {\bibinfo
  {volume} {151}},\ \bibinfo {pages} {101} (\bibinfo {year} {2022})},\ \Eprint
  {https://arxiv.org/abs/2305.10032} {arXiv:2305.10032} \BibitemShut {NoStop}%
\bibitem [{\citenamefont {Spirtes}\ \emph {et~al.}(1995)\citenamefont
  {Spirtes}, \citenamefont {Meek},\ and\ \citenamefont
  {Richardson}}]{spirtes1995}%
  \BibitemOpen
  \bibfield  {author} {\bibinfo {author} {\bibfnamefont {P.}~\bibnamefont
  {Spirtes}}, \bibinfo {author} {\bibfnamefont {C.}~\bibnamefont {Meek}},\ and\
  \bibinfo {author} {\bibfnamefont {T.}~\bibnamefont {Richardson}},\ }\bibfield
   {title} {\bibinfo {title} {Causal inference in the presence of latent
  variables and selection bias},\ }in\ \href@noop {} {\emph {\bibinfo
  {booktitle} {Proc. 11th Conf. Uncertain. Artif. Intell. (UAI1995)}}},\
  \bibinfo {series and number} {UAI'95}\ (\bibinfo  {publisher} {Morgan
  Kaufmann Publishers Inc.},\ \bibinfo {address} {San Francisco, CA, USA},\
  \bibinfo {year} {1995})\ p.\ \bibinfo {pages} {499–506},\ \Eprint
  {https://arxiv.org/abs/1302.4983} {arXiv:1302.4983} \BibitemShut {NoStop}%
\bibitem [{\citenamefont {Colombo}\ and\ \citenamefont
  {Maathuis}(2014)}]{colombo2014}%
  \BibitemOpen
  \bibfield  {author} {\bibinfo {author} {\bibfnamefont {D.}~\bibnamefont
  {Colombo}}\ and\ \bibinfo {author} {\bibfnamefont {M.~H.}\ \bibnamefont
  {Maathuis}},\ }\bibfield  {title} {\bibinfo {title} {Order-independent
  constraint-based causal structure learning},\ }\href
  {http://jmlr.org/papers/v15/colombo14a.html} {\bibfield  {journal} {\bibinfo
  {journal} {J. Mach. Learn. Res.}\ }\textbf {\bibinfo {volume} {15}},\
  \bibinfo {pages} {3921} (\bibinfo {year} {2014})},\ \Eprint
  {https://arxiv.org/abs/1211.3295} {arXiv:1211.3295} \BibitemShut {NoStop}%
\bibitem [{\citenamefont {Le}\ \emph {et~al.}(2019)\citenamefont {Le},
  \citenamefont {Hoang}, \citenamefont {Li}, \citenamefont {Liu}, \citenamefont
  {Liu},\ and\ \citenamefont {Hu}}]{le2019}%
  \BibitemOpen
  \bibfield  {author} {\bibinfo {author} {\bibfnamefont {T.~D.}\ \bibnamefont
  {Le}}, \bibinfo {author} {\bibfnamefont {T.}~\bibnamefont {Hoang}}, \bibinfo
  {author} {\bibfnamefont {J.}~\bibnamefont {Li}}, \bibinfo {author}
  {\bibfnamefont {L.}~\bibnamefont {Liu}}, \bibinfo {author} {\bibfnamefont
  {H.}~\bibnamefont {Liu}},\ and\ \bibinfo {author} {\bibfnamefont
  {S.}~\bibnamefont {Hu}},\ }\bibfield  {title} {\bibinfo {title} {A fast {PC}
  algorithm for high dimensional causal discovery with multi-core {PCs}},\
  }\href {https://doi.org/10.1109/TCBB.2016.2591526} {\bibfield  {journal}
  {\bibinfo  {journal} {IEEE/ACM Trans. Comput. Biology Bioinform.}\ }\textbf
  {\bibinfo {volume} {16}},\ \bibinfo {pages} {1483} (\bibinfo {year}
  {2019})},\ \Eprint {https://arxiv.org/abs/1502.02454} {arXiv:1502.02454}
  \BibitemShut {NoStop}%
\bibitem [{\citenamefont {Pasquato}\ \emph {et~al.}(2023)\citenamefont
  {Pasquato}, \citenamefont {Jin}, \citenamefont {Lemos}, \citenamefont
  {Davis},\ and\ \citenamefont {Maccio}}]{pasquato2023}%
  \BibitemOpen
  \bibfield  {author} {\bibinfo {author} {\bibfnamefont {M.}~\bibnamefont
  {Pasquato}}, \bibinfo {author} {\bibfnamefont {Z.}~\bibnamefont {Jin}},
  \bibinfo {author} {\bibfnamefont {P.}~\bibnamefont {Lemos}}, \bibinfo
  {author} {\bibfnamefont {B.~L.}\ \bibnamefont {Davis}},\ and\ \bibinfo
  {author} {\bibfnamefont {A.}~\bibnamefont {Maccio}},\ }\bibfield  {title}
  {\bibinfo {title} {Causa prima: cosmology meets causal discovery for the
  first time},\ }in\ \href
  {https://ml4physicalsciences.github.io/2023/files/NeurIPS_ML4PS_2023_62.pdf}
  {\emph {\bibinfo {booktitle} {NeurIPS 2023 Machine Learning and the Physical
  Sciences Workshop}}}\ (\bibinfo {year} {2023})\ \Eprint
  {https://arxiv.org/abs/2311.15160} {arXiv:2311.15160} \BibitemShut {NoStop}%
\bibitem [{\citenamefont {Akaike}(1974)}]{akaike1974}%
  \BibitemOpen
  \bibfield  {author} {\bibinfo {author} {\bibfnamefont {H.}~\bibnamefont
  {Akaike}},\ }\bibfield  {title} {\bibinfo {title} {A new look at the
  statistical model identification},\ }\href
  {https://doi.org/10.1109/TAC.1974.1100705} {\bibfield  {journal} {\bibinfo
  {journal} {IEEE Trans. Autom. Control}\ }\textbf {\bibinfo {volume} {19}},\
  \bibinfo {pages} {716} (\bibinfo {year} {1974})}\BibitemShut {NoStop}%
\bibitem [{\citenamefont {Geiger}\ and\ \citenamefont
  {Heckerman}(1994)}]{geiger1994}%
  \BibitemOpen
  \bibfield  {author} {\bibinfo {author} {\bibfnamefont {D.}~\bibnamefont
  {Geiger}}\ and\ \bibinfo {author} {\bibfnamefont {D.}~\bibnamefont
  {Heckerman}},\ }\bibfield  {title} {\bibinfo {title} {Learning {Gaussian}
  networks},\ }in\ \href
  {https://doi.org/https://doi.org/10.1016/B978-1-55860-332-5.50035-3} {\emph
  {\bibinfo {booktitle} {Uncertainty in Artificial Intelligence}}},\ \bibinfo
  {editor} {edited by\ \bibinfo {editor} {\bibfnamefont {R.~L.}\ \bibnamefont
  {{de Mantaras}}}\ and\ \bibinfo {editor} {\bibfnamefont {D.}~\bibnamefont
  {Poole}}}\ (\bibinfo  {publisher} {Morgan Kaufmann},\ \bibinfo {address} {San
  Francisco (CA)},\ \bibinfo {year} {1994})\ pp.\ \bibinfo {pages} {235--243},\
  \Eprint {https://arxiv.org/abs/1302.6808} {arXiv:1302.6808} \BibitemShut
  {NoStop}%
\bibitem [{\citenamefont {Jin}\ \emph {et~al.}(2024)\citenamefont {Jin},
  \citenamefont {Pasquato}, \citenamefont {Davis}, \citenamefont {Maccio},\
  and\ \citenamefont {Hezaveh}}]{jin2024}%
  \BibitemOpen
  \bibfield  {author} {\bibinfo {author} {\bibfnamefont {Z.}~\bibnamefont
  {Jin}}, \bibinfo {author} {\bibfnamefont {M.}~\bibnamefont {Pasquato}},
  \bibinfo {author} {\bibfnamefont {B.~L.}\ \bibnamefont {Davis}}, \bibinfo
  {author} {\bibfnamefont {A.}~\bibnamefont {Maccio}},\ and\ \bibinfo {author}
  {\bibfnamefont {Y.}~\bibnamefont {Hezaveh}},\ }\bibfield  {title} {\bibinfo
  {title} {Beyond causal discovery for astronomy: Learning meaningful
  representations with independent component analysis},\ }in\ \href
  {https://openreview.net/forum?id=Bygf3JeFDb} {\emph {\bibinfo {booktitle}
  {NeurIPS 2024 Causal Representation Learning Workshop}}}\ (\bibinfo {year}
  {2024})\ \Eprint {https://arxiv.org/abs/2410.14775} {arXiv:2410.14775}
  \BibitemShut {NoStop}%
\bibitem [{\citenamefont {Gross}(1996)}]{gross1996}%
  \BibitemOpen
  \bibfield  {author} {\bibinfo {author} {\bibfnamefont {D.~J.}\ \bibnamefont
  {Gross}},\ }\bibfield  {title} {\bibinfo {title} {The role of symmetry in
  fundamental physics},\ }\href {https://doi.org/10.1073/pnas.93.25.14256}
  {\bibfield  {journal} {\bibinfo  {journal} {Proc. Natl. Acad. Sci.}\ }\textbf
  {\bibinfo {volume} {93}},\ \bibinfo {pages} {14256} (\bibinfo {year}
  {1996})}\BibitemShut {NoStop}%
\bibitem [{\citenamefont {Bell}(1964)}]{bell1964}%
  \BibitemOpen
  \bibfield  {author} {\bibinfo {author} {\bibfnamefont {J.~S.}\ \bibnamefont
  {Bell}},\ }\bibfield  {title} {\bibinfo {title} {On the {E}instein {P}odolsky
  {R}osen paradox},\ }\href
  {https://doi.org/10.1103/PhysicsPhysiqueFizika.1.195} {\bibfield  {journal}
  {\bibinfo  {journal} {Physics Physique Fizika}\ }\textbf {\bibinfo {volume}
  {1}},\ \bibinfo {pages} {195} (\bibinfo {year} {1964})}\BibitemShut {NoStop}%
\bibitem [{\citenamefont {Tavakoli}\ \emph {et~al.}(2022)\citenamefont
  {Tavakoli}, \citenamefont {Pozas-Kerstjens}, \citenamefont {Luo},\ and\
  \citenamefont {Renou}}]{TavakoliPozas2022}%
  \BibitemOpen
  \bibfield  {author} {\bibinfo {author} {\bibfnamefont {A.}~\bibnamefont
  {Tavakoli}}, \bibinfo {author} {\bibfnamefont {A.}~\bibnamefont
  {Pozas-Kerstjens}}, \bibinfo {author} {\bibfnamefont {M.-X.}\ \bibnamefont
  {Luo}},\ and\ \bibinfo {author} {\bibfnamefont {M.-O.}\ \bibnamefont
  {Renou}},\ }\bibfield  {title} {\bibinfo {title} {Bell nonlocality in
  networks},\ }\href {https://doi.org/10.1088/1361-6633/ac41bb} {\bibfield
  {journal} {\bibinfo  {journal} {Rep. Prog. Phys.}\ }\textbf {\bibinfo
  {volume} {85}},\ \bibinfo {pages} {056001} (\bibinfo {year} {2022})},\
  \Eprint {https://arxiv.org/abs/2104.10700} {arXiv:2104.10700} \BibitemShut
  {NoStop}%
\bibitem [{\citenamefont {Fritz}(2012)}]{fritz2012}%
  \BibitemOpen
  \bibfield  {author} {\bibinfo {author} {\bibfnamefont {T.}~\bibnamefont
  {Fritz}},\ }\bibfield  {title} {\bibinfo {title} {Beyond {Bell}'s theorem:
  correlation scenarios},\ }\href
  {https://doi.org/10.1088/1367-2630/14/10/103001} {\bibfield  {journal}
  {\bibinfo  {journal} {New J. Phys.}\ }\textbf {\bibinfo {volume} {14}},\
  \bibinfo {pages} {103001} (\bibinfo {year} {2012})},\ \Eprint
  {https://arxiv.org/abs/1206.5115} {arXiv:1206.5115} \BibitemShut {NoStop}%
\bibitem [{\citenamefont {Renou}\ \emph {et~al.}(2019)\citenamefont {Renou},
  \citenamefont {B\"aumer}, \citenamefont {Boreiri}, \citenamefont {Brunner},
  \citenamefont {Gisin},\ and\ \citenamefont {Beigi}}]{renou2019}%
  \BibitemOpen
  \bibfield  {author} {\bibinfo {author} {\bibfnamefont {M.-O.}\ \bibnamefont
  {Renou}}, \bibinfo {author} {\bibfnamefont {E.}~\bibnamefont {B\"aumer}},
  \bibinfo {author} {\bibfnamefont {S.}~\bibnamefont {Boreiri}}, \bibinfo
  {author} {\bibfnamefont {N.}~\bibnamefont {Brunner}}, \bibinfo {author}
  {\bibfnamefont {N.}~\bibnamefont {Gisin}},\ and\ \bibinfo {author}
  {\bibfnamefont {S.}~\bibnamefont {Beigi}},\ }\bibfield  {title} {\bibinfo
  {title} {Genuine quantum nonlocality in the triangle network},\ }\href
  {https://doi.org/10.1103/PhysRevLett.123.140401} {\bibfield  {journal}
  {\bibinfo  {journal} {Phys. Rev. Lett.}\ }\textbf {\bibinfo {volume} {123}},\
  \bibinfo {pages} {140401} (\bibinfo {year} {2019})},\ \Eprint
  {https://arxiv.org/abs/1905.04902} {arXiv:1905.04902} \BibitemShut {NoStop}%
\bibitem [{\citenamefont {Fraser}\ and\ \citenamefont
  {Wolfe}(2018)}]{fraser2018}%
  \BibitemOpen
  \bibfield  {author} {\bibinfo {author} {\bibfnamefont {T.~C.}\ \bibnamefont
  {Fraser}}\ and\ \bibinfo {author} {\bibfnamefont {E.}~\bibnamefont {Wolfe}},\
  }\bibfield  {title} {\bibinfo {title} {Causal compatibility inequalities
  admitting quantum violations in the triangle structure},\ }\href
  {https://doi.org/10.1103/PhysRevA.98.022113} {\bibfield  {journal} {\bibinfo
  {journal} {Phys. Rev. A}\ }\textbf {\bibinfo {volume} {98}},\ \bibinfo
  {pages} {022113} (\bibinfo {year} {2018})},\ \Eprint
  {https://arxiv.org/abs/1709.06242} {arXiv:1709.06242} \BibitemShut {NoStop}%
\bibitem [{\citenamefont {Gisin}\ \emph {et~al.}(2020)\citenamefont {Gisin},
  \citenamefont {Bancal}, \citenamefont {Cai}, \citenamefont {Remy},
  \citenamefont {Tavakoli}, \citenamefont {Zambrini~Cruzeiro}, \citenamefont
  {Popescu},\ and\ \citenamefont {Brunner}}]{gisin2020}%
  \BibitemOpen
  \bibfield  {author} {\bibinfo {author} {\bibfnamefont {N.}~\bibnamefont
  {Gisin}}, \bibinfo {author} {\bibfnamefont {J.-D.}\ \bibnamefont {Bancal}},
  \bibinfo {author} {\bibfnamefont {Y.}~\bibnamefont {Cai}}, \bibinfo {author}
  {\bibfnamefont {P.}~\bibnamefont {Remy}}, \bibinfo {author} {\bibfnamefont
  {A.}~\bibnamefont {Tavakoli}}, \bibinfo {author} {\bibfnamefont
  {E.}~\bibnamefont {Zambrini~Cruzeiro}}, \bibinfo {author} {\bibfnamefont
  {S.}~\bibnamefont {Popescu}},\ and\ \bibinfo {author} {\bibfnamefont
  {N.}~\bibnamefont {Brunner}},\ }\bibfield  {title} {\bibinfo {title}
  {Constraints on nonlocality in networks from no-signaling and independence},\
  }\href {https://doi.org/10.1038/s41467-020-16137-4} {\bibfield  {journal}
  {\bibinfo  {journal} {Nat. Commun.}\ }\textbf {\bibinfo {volume} {11}},\
  \bibinfo {pages} {2378} (\bibinfo {year} {2020})},\ \Eprint
  {https://arxiv.org/abs/1906.06495} {arXiv:1906.06495} \BibitemShut {NoStop}%
\bibitem [{\citenamefont {Pozas-Kerstjens}\ \emph
  {et~al.}(2023{\natexlab{a}})\citenamefont {Pozas-Kerstjens}, \citenamefont
  {Girardin}, \citenamefont {Kriváchy}, \citenamefont {Tavakoli},\ and\
  \citenamefont {Gisin}}]{pozas2023}%
  \BibitemOpen
  \bibfield  {author} {\bibinfo {author} {\bibfnamefont {A.}~\bibnamefont
  {Pozas-Kerstjens}}, \bibinfo {author} {\bibfnamefont {A.}~\bibnamefont
  {Girardin}}, \bibinfo {author} {\bibfnamefont {T.}~\bibnamefont {Kriváchy}},
  \bibinfo {author} {\bibfnamefont {A.}~\bibnamefont {Tavakoli}},\ and\
  \bibinfo {author} {\bibfnamefont {N.}~\bibnamefont {Gisin}},\ }\bibfield
  {title} {\bibinfo {title} {Post-quantum nonlocality in the minimal triangle
  scenario},\ }\href {https://doi.org/10.1088/1367-2630/ad0a16} {\bibfield
  {journal} {\bibinfo  {journal} {New J. Phys.}\ }\textbf {\bibinfo {volume}
  {25}},\ \bibinfo {pages} {113037} (\bibinfo {year} {2023}{\natexlab{a}})},\
  \Eprint {https://arxiv.org/abs/2305.03745} {arXiv:2305.03745} \BibitemShut
  {NoStop}%
\bibitem [{\citenamefont {Wood}\ and\ \citenamefont
  {Spekkens}(2015)}]{wood2015}%
  \BibitemOpen
  \bibfield  {author} {\bibinfo {author} {\bibfnamefont {C.~J.}\ \bibnamefont
  {Wood}}\ and\ \bibinfo {author} {\bibfnamefont {R.~W.}\ \bibnamefont
  {Spekkens}},\ }\bibfield  {title} {\bibinfo {title} {The lesson of causal
  discovery algorithms for quantum correlations: causal explanations of
  {B}ell-inequality violations require fine-tuning},\ }\href
  {https://doi.org/10.1088/1367-2630/17/3/033002} {\bibfield  {journal}
  {\bibinfo  {journal} {New J. Phys.}\ }\textbf {\bibinfo {volume} {17}},\
  \bibinfo {pages} {033002} (\bibinfo {year} {2015})},\ \Eprint
  {https://arxiv.org/abs/1208.4119} {arXiv:1208.4119} \BibitemShut {NoStop}%
\bibitem [{\citenamefont {Henson}\ \emph {et~al.}(2014)\citenamefont {Henson},
  \citenamefont {Lal},\ and\ \citenamefont {Pusey}}]{Henson2014}%
  \BibitemOpen
  \bibfield  {author} {\bibinfo {author} {\bibfnamefont {J.}~\bibnamefont
  {Henson}}, \bibinfo {author} {\bibfnamefont {R.}~\bibnamefont {Lal}},\ and\
  \bibinfo {author} {\bibfnamefont {M.~F.}\ \bibnamefont {Pusey}},\ }\bibfield
  {title} {\bibinfo {title} {Theory-independent limits on correlations from
  generalized {B}ayesian networks},\ }\href
  {https://doi.org/10.1088/1367-2630/16/11/113043} {\bibfield  {journal}
  {\bibinfo  {journal} {New J. Phys.}\ }\textbf {\bibinfo {volume} {16}},\
  \bibinfo {pages} {113043} (\bibinfo {year} {2014})},\ \Eprint
  {https://arxiv.org/abs/1405.2572} {arXiv:1405.2572} \BibitemShut {NoStop}%
\bibitem [{\citenamefont {Weilenmann}\ and\ \citenamefont
  {Colbeck}(2020)}]{Weilenmann2020}%
  \BibitemOpen
  \bibfield  {author} {\bibinfo {author} {\bibfnamefont {M.}~\bibnamefont
  {Weilenmann}}\ and\ \bibinfo {author} {\bibfnamefont {R.}~\bibnamefont
  {Colbeck}},\ }\bibfield  {title} {\bibinfo {title} {Analysing causal
  structures in generalised probabilistic theories},\ }\href
  {https://doi.org/10.22331/q-2020-02-27-236} {\bibfield  {journal} {\bibinfo
  {journal} {Quantum}\ }\textbf {\bibinfo {volume} {4}},\ \bibinfo {pages}
  {236} (\bibinfo {year} {2020})},\ \Eprint {https://arxiv.org/abs/1812.04327}
  {arXiv:1812.04327} \BibitemShut {NoStop}%
\bibitem [{\citenamefont {Janotta}\ and\ \citenamefont
  {Hinrichsen}(2014)}]{janotta2014}%
  \BibitemOpen
  \bibfield  {author} {\bibinfo {author} {\bibfnamefont {P.}~\bibnamefont
  {Janotta}}\ and\ \bibinfo {author} {\bibfnamefont {H.}~\bibnamefont
  {Hinrichsen}},\ }\bibfield  {title} {\bibinfo {title} {Generalized
  probability theories: what determines the structure of quantum theory?},\
  }\href {https://doi.org/10.1088/1751-8113/47/32/323001} {\bibfield  {journal}
  {\bibinfo  {journal} {J. Phys. A: Math. Theor.}\ }\textbf {\bibinfo {volume}
  {47}},\ \bibinfo {pages} {323001} (\bibinfo {year} {2014})},\ \Eprint
  {https://arxiv.org/abs/1402.6562} {arXiv:1402.6562} \BibitemShut {NoStop}%
\bibitem [{\citenamefont {Plávala}(2023)}]{plavala2023}%
  \BibitemOpen
  \bibfield  {author} {\bibinfo {author} {\bibfnamefont {M.}~\bibnamefont
  {Plávala}},\ }\bibfield  {title} {\bibinfo {title} {General probabilistic
  theories: An introduction},\ }\href
  {https://doi.org/https://doi.org/10.1016/j.physrep.2023.09.001} {\bibfield
  {journal} {\bibinfo  {journal} {Phys. Rep.}\ }\textbf {\bibinfo {volume}
  {1033}},\ \bibinfo {pages} {1} (\bibinfo {year} {2023})},\ \Eprint
  {https://arxiv.org/abs/2103.07469} {arXiv:2103.07469} \BibitemShut {NoStop}%
\bibitem [{\citenamefont {Ludwig}(1983)}]{ludwigbook}%
  \BibitemOpen
  \bibfield  {author} {\bibinfo {author} {\bibfnamefont {G.}~\bibnamefont
  {Ludwig}},\ }\href {https://web.stanford.edu/~boyd/cvxbook/} {\emph {\bibinfo
  {title} {Foundations of Quantum Mechanics}}},\ Texts and Monographs in
  Physics\ (\bibinfo  {publisher} {{Springer}},\ \bibinfo {year}
  {1983})\BibitemShut {NoStop}%
\bibitem [{\citenamefont {Kraus}(1983)}]{krausbook}%
  \BibitemOpen
  \bibfield  {author} {\bibinfo {author} {\bibfnamefont {K.}~\bibnamefont
  {Kraus}},\ }\href {https://doi.org/10.1007/3-540-12732-1} {\emph {\bibinfo
  {title} {States, Effects, and Operations. Fundamental Notions of Quantum
  Theory}}},\ edited by\ \bibinfo {editor} {\bibfnamefont {K.}~\bibnamefont
  {Kraus}}, \bibinfo {editor} {\bibfnamefont {A.}~\bibnamefont {Böhm}},
  \bibinfo {editor} {\bibfnamefont {J.~D.}\ \bibnamefont {Dollard}},\ and\
  \bibinfo {editor} {\bibfnamefont {W.~H.}\ \bibnamefont {Wootters}},\ \bibinfo
  {series} {Lecture Notes in Physics}, Vol.\ \bibinfo {volume} {190}\ (\bibinfo
   {publisher} {{Springer Berlin, Heidelberg}},\ \bibinfo {year}
  {1983})\BibitemShut {NoStop}%
\bibitem [{\citenamefont {da~Silva}\ and\ \citenamefont
  {Parisio}(2023)}]{silva2023}%
  \BibitemOpen
  \bibfield  {author} {\bibinfo {author} {\bibfnamefont {J.~M.}\ \bibnamefont
  {da~Silva}}\ and\ \bibinfo {author} {\bibfnamefont {F.}~\bibnamefont
  {Parisio}},\ }\bibfield  {title} {\bibinfo {title} {Numerically assisted
  determination of local models in network scenarios},\ }\href
  {https://doi.org/10.1103/PhysRevA.108.052602} {\bibfield  {journal} {\bibinfo
   {journal} {Phys. Rev. A}\ }\textbf {\bibinfo {volume} {108}},\ \bibinfo
  {pages} {052602} (\bibinfo {year} {2023})},\ \Eprint
  {https://arxiv.org/abs/2303.09954} {arXiv:2303.09954} \BibitemShut {NoStop}%
\bibitem [{\citenamefont {Boyd}\ and\ \citenamefont
  {Vandenberghe}(2004)}]{boydbook}%
  \BibitemOpen
  \bibfield  {author} {\bibinfo {author} {\bibfnamefont {S.}~\bibnamefont
  {Boyd}}\ and\ \bibinfo {author} {\bibfnamefont {L.}~\bibnamefont
  {Vandenberghe}},\ }\href {https://web.stanford.edu/~boyd/cvxbook/} {\emph
  {\bibinfo {title} {Convex Optimization}}}\ (\bibinfo  {publisher} {{Cambridge
  University Press}},\ \bibinfo {address} {Cambridge, England},\ \bibinfo
  {year} {2004})\BibitemShut {NoStop}%
\bibitem [{\citenamefont {Pozas-Kerstjens}(2024)}]{compapp}%
  \BibitemOpen
  \bibfield  {author} {\bibinfo {author} {\bibfnamefont {A.}~\bibnamefont
  {Pozas-Kerstjens}},\ }\href {https://doi.org/10.5281/zenodo.14899767}
  {}\bibinfo {howpublished}
  {\href{https://www.doi.org/10.5281/zenodo.14899767}{Zenodo 14899767}}
  (\bibinfo {year} {2024})\BibitemShut {NoStop}%
\bibitem [{\citenamefont {Wolfe}\ \emph {et~al.}(2019)\citenamefont {Wolfe},
  \citenamefont {Spekkens},\ and\ \citenamefont {Fritz}}]{wolfe2019}%
  \BibitemOpen
  \bibfield  {author} {\bibinfo {author} {\bibfnamefont {E.}~\bibnamefont
  {Wolfe}}, \bibinfo {author} {\bibfnamefont {R.~W.}\ \bibnamefont
  {Spekkens}},\ and\ \bibinfo {author} {\bibfnamefont {T.}~\bibnamefont
  {Fritz}},\ }\bibfield  {title} {\bibinfo {title} {The inflation technique for
  causal inference with latent variables},\ }\href
  {https://doi.org/10.1515/jci-2017-0020} {\bibfield  {journal} {\bibinfo
  {journal} {J. Causal Inference}\ }\textbf {\bibinfo {volume} {7}},\ \bibinfo
  {pages} {20170020} (\bibinfo {year} {2019})},\ \Eprint
  {https://arxiv.org/abs/1609.00672} {arXiv:1609.00672} \BibitemShut {NoStop}%
\bibitem [{\citenamefont {Pozas-Kerstjens}\ \emph
  {et~al.}(2023{\natexlab{b}})\citenamefont {Pozas-Kerstjens}, \citenamefont
  {Gisin},\ and\ \citenamefont {Renou}}]{pozas2022b}%
  \BibitemOpen
  \bibfield  {author} {\bibinfo {author} {\bibfnamefont {A.}~\bibnamefont
  {Pozas-Kerstjens}}, \bibinfo {author} {\bibfnamefont {N.}~\bibnamefont
  {Gisin}},\ and\ \bibinfo {author} {\bibfnamefont {M.-O.}\ \bibnamefont
  {Renou}},\ }\bibfield  {title} {\bibinfo {title} {Proofs of network quantum
  nonlocality in continuous families of distributions},\ }\href
  {https://doi.org/10.1103/PhysRevLett.130.090201} {\bibfield  {journal}
  {\bibinfo  {journal} {Phys. Rev. Lett.}\ }\textbf {\bibinfo {volume} {130}},\
  \bibinfo {pages} {090201} (\bibinfo {year} {2023}{\natexlab{b}})},\ \Eprint
  {https://arxiv.org/abs/2203.16543} {arXiv:2203.16543} \BibitemShut {NoStop}%
\bibitem [{\citenamefont {Lauand}\ \emph {et~al.}(2024)\citenamefont {Lauand},
  \citenamefont {Bekele},\ and\ \citenamefont {Wolfe}}]{lauand2024}%
  \BibitemOpen
  \bibfield  {author} {\bibinfo {author} {\bibfnamefont {P.}~\bibnamefont
  {Lauand}}, \bibinfo {author} {\bibfnamefont {B.~N.}\ \bibnamefont {Bekele}},\
  and\ \bibinfo {author} {\bibfnamefont {E.}~\bibnamefont {Wolfe}},\
  }\href@noop {} {\bibinfo {title} {Quantum non-classicality from causal data
  fusion}} (\bibinfo {year} {2024}),\ \Eprint
  {https://arxiv.org/abs/2405.19252} {arXiv:2405.19252} \BibitemShut {NoStop}%
\bibitem [{\citenamefont {Lauand}\ \emph {et~al.}(2023)\citenamefont {Lauand},
  \citenamefont {Poderini}, \citenamefont {Nery}, \citenamefont {Moreno},
  \citenamefont {Pollyceno}, \citenamefont {Rabelo},\ and\ \citenamefont
  {Chaves}}]{lauand2023}%
  \BibitemOpen
  \bibfield  {author} {\bibinfo {author} {\bibfnamefont {P.}~\bibnamefont
  {Lauand}}, \bibinfo {author} {\bibfnamefont {D.}~\bibnamefont {Poderini}},
  \bibinfo {author} {\bibfnamefont {R.}~\bibnamefont {Nery}}, \bibinfo {author}
  {\bibfnamefont {G.}~\bibnamefont {Moreno}}, \bibinfo {author} {\bibfnamefont
  {L.}~\bibnamefont {Pollyceno}}, \bibinfo {author} {\bibfnamefont
  {R.}~\bibnamefont {Rabelo}},\ and\ \bibinfo {author} {\bibfnamefont
  {R.}~\bibnamefont {Chaves}},\ }\bibfield  {title} {\bibinfo {title}
  {Witnessing nonclassicality in a causal structure with three observable
  variables},\ }\href {https://doi.org/10.1103/PRXQuantum.4.020311} {\bibfield
  {journal} {\bibinfo  {journal} {PRX Quantum}\ }\textbf {\bibinfo {volume}
  {4}},\ \bibinfo {pages} {020311} (\bibinfo {year} {2023})},\ \Eprint
  {https://arxiv.org/abs/2211.13349} {arXiv:2211.13349} \BibitemShut {NoStop}%
\bibitem [{\citenamefont {Wolfe}\ \emph {et~al.}(2021)\citenamefont {Wolfe},
  \citenamefont {Pozas-Kerstjens}, \citenamefont {Grinberg}, \citenamefont
  {Rosset}, \citenamefont {Ac\'{\i}n},\ and\ \citenamefont
  {Navascu\'es}}]{wolfe2021}%
  \BibitemOpen
  \bibfield  {author} {\bibinfo {author} {\bibfnamefont {E.}~\bibnamefont
  {Wolfe}}, \bibinfo {author} {\bibfnamefont {A.}~\bibnamefont
  {Pozas-Kerstjens}}, \bibinfo {author} {\bibfnamefont {M.}~\bibnamefont
  {Grinberg}}, \bibinfo {author} {\bibfnamefont {D.}~\bibnamefont {Rosset}},
  \bibinfo {author} {\bibfnamefont {A.}~\bibnamefont {Ac\'{\i}n}},\ and\
  \bibinfo {author} {\bibfnamefont {M.}~\bibnamefont {Navascu\'es}},\
  }\bibfield  {title} {\bibinfo {title} {Quantum inflation: A general approach
  to quantum causal compatibility},\ }\href
  {https://doi.org/10.1103/PhysRevX.11.021043} {\bibfield  {journal} {\bibinfo
  {journal} {Phys. Rev. X}\ }\textbf {\bibinfo {volume} {11}},\ \bibinfo
  {pages} {021043} (\bibinfo {year} {2021})},\ \Eprint
  {https://arxiv.org/abs/1909.10519} {arXiv:1909.10519} \BibitemShut {NoStop}%
\bibitem [{\citenamefont {Camillo}\ \emph {et~al.}(2024)\citenamefont
  {Camillo}, \citenamefont {Lauand}, \citenamefont {Poderini}, \citenamefont
  {Rabelo},\ and\ \citenamefont {Chaves}}]{camillo2023}%
  \BibitemOpen
  \bibfield  {author} {\bibinfo {author} {\bibfnamefont {G.}~\bibnamefont
  {Camillo}}, \bibinfo {author} {\bibfnamefont {P.}~\bibnamefont {Lauand}},
  \bibinfo {author} {\bibfnamefont {D.}~\bibnamefont {Poderini}}, \bibinfo
  {author} {\bibfnamefont {R.}~\bibnamefont {Rabelo}},\ and\ \bibinfo {author}
  {\bibfnamefont {R.}~\bibnamefont {Chaves}},\ }\bibfield  {title} {\bibinfo
  {title} {Estimating the volume of correlation sets in causal networks},\
  }\href {https://doi.org/10.1103/PhysRevA.109.012220} {\bibfield  {journal}
  {\bibinfo  {journal} {Phys. Rev. A}\ }\textbf {\bibinfo {volume} {109}},\
  \bibinfo {pages} {012220} (\bibinfo {year} {2024})},\ \Eprint
  {https://arxiv.org/abs/2311.08574} {arXiv:2311.08574} \BibitemShut {NoStop}%
\bibitem [{\citenamefont {Pozas-Kerstjens}\ \emph {et~al.}(2022)\citenamefont
  {Pozas-Kerstjens}, \citenamefont {Gisin},\ and\ \citenamefont
  {Tavakoli}}]{pozas2022}%
  \BibitemOpen
  \bibfield  {author} {\bibinfo {author} {\bibfnamefont {A.}~\bibnamefont
  {Pozas-Kerstjens}}, \bibinfo {author} {\bibfnamefont {N.}~\bibnamefont
  {Gisin}},\ and\ \bibinfo {author} {\bibfnamefont {A.}~\bibnamefont
  {Tavakoli}},\ }\bibfield  {title} {\bibinfo {title} {Full network
  nonlocality},\ }\href {https://doi.org/10.1103/PhysRevLett.128.010403}
  {\bibfield  {journal} {\bibinfo  {journal} {Phys. Rev. Lett.}\ }\textbf
  {\bibinfo {volume} {128}},\ \bibinfo {pages} {010403} (\bibinfo {year}
  {2022})},\ \Eprint {https://arxiv.org/abs/2105.09325} {arXiv:2105.09325}
  \BibitemShut {NoStop}%
\bibitem [{\citenamefont {Wang}\ \emph {et~al.}(2023)\citenamefont {Wang},
  \citenamefont {Pozas-Kerstjens}, \citenamefont {Zhang}, \citenamefont {Liu},
  \citenamefont {Huang}, \citenamefont {Li}, \citenamefont {Guo}, \citenamefont
  {Gisin},\ and\ \citenamefont {Tavakoli}}]{wang2023}%
  \BibitemOpen
  \bibfield  {author} {\bibinfo {author} {\bibfnamefont {N.-N.}\ \bibnamefont
  {Wang}}, \bibinfo {author} {\bibfnamefont {A.}~\bibnamefont
  {Pozas-Kerstjens}}, \bibinfo {author} {\bibfnamefont {C.}~\bibnamefont
  {Zhang}}, \bibinfo {author} {\bibfnamefont {B.-H.}\ \bibnamefont {Liu}},
  \bibinfo {author} {\bibfnamefont {Y.-F.}\ \bibnamefont {Huang}}, \bibinfo
  {author} {\bibfnamefont {C.-F.}\ \bibnamefont {Li}}, \bibinfo {author}
  {\bibfnamefont {G.-C.}\ \bibnamefont {Guo}}, \bibinfo {author} {\bibfnamefont
  {N.}~\bibnamefont {Gisin}},\ and\ \bibinfo {author} {\bibfnamefont
  {A.}~\bibnamefont {Tavakoli}},\ }\bibfield  {title} {\bibinfo {title}
  {Certification of non-classicality in all links of a photonic star network
  without assuming quantum mechanics},\ }\href
  {https://doi.org/10.1038/s41467-023-37842-w} {\bibfield  {journal} {\bibinfo
  {journal} {Nat. Commun.}\ }\textbf {\bibinfo {volume} {14}},\ \bibinfo
  {pages} {2153} (\bibinfo {year} {2023})},\ \Eprint
  {https://arxiv.org/abs/2212.09765} {arXiv:2212.09765} \BibitemShut {NoStop}%
\bibitem [{\citenamefont {Tavakoli}\ \emph {et~al.}(2024)\citenamefont
  {Tavakoli}, \citenamefont {Pozas-Kerstjens}, \citenamefont {Brown},\ and\
  \citenamefont {Ara\'ujo}}]{TavakoliPozas2024}%
  \BibitemOpen
  \bibfield  {author} {\bibinfo {author} {\bibfnamefont {A.}~\bibnamefont
  {Tavakoli}}, \bibinfo {author} {\bibfnamefont {A.}~\bibnamefont
  {Pozas-Kerstjens}}, \bibinfo {author} {\bibfnamefont {P.}~\bibnamefont
  {Brown}},\ and\ \bibinfo {author} {\bibfnamefont {M.}~\bibnamefont
  {Ara\'ujo}},\ }\bibfield  {title} {\bibinfo {title} {Semidefinite programming
  relaxations for quantum correlations},\ }\href
  {https://doi.org/10.1103/RevModPhys.96.045006} {\bibfield  {journal}
  {\bibinfo  {journal} {Rev. Mod. Phys.}\ }\textbf {\bibinfo {volume} {96}},\
  \bibinfo {pages} {045006} (\bibinfo {year} {2024})},\ \Eprint
  {https://arxiv.org/abs/2307.02551} {arXiv:2307.02551} \BibitemShut {NoStop}%
\bibitem [{\citenamefont {Boghiu}\ \emph {et~al.}(2023)\citenamefont {Boghiu},
  \citenamefont {Wolfe},\ and\ \citenamefont {Pozas-Kerstjens}}]{boghiu2022}%
  \BibitemOpen
  \bibfield  {author} {\bibinfo {author} {\bibfnamefont {E.-C.}\ \bibnamefont
  {Boghiu}}, \bibinfo {author} {\bibfnamefont {E.}~\bibnamefont {Wolfe}},\ and\
  \bibinfo {author} {\bibfnamefont {A.}~\bibnamefont {Pozas-Kerstjens}},\
  }\bibfield  {title} {\bibinfo {title} {Inflation: a {P}ython library for
  classical and quantum causal compatibility},\ }\href
  {https://doi.org/10.22331/q-2023-05-04-996} {\bibfield  {journal} {\bibinfo
  {journal} {{Quantum}}\ }\textbf {\bibinfo {volume} {7}},\ \bibinfo {pages}
  {996} (\bibinfo {year} {2023})},\ \bibinfo {note}
  {\url{https://github.com/ecboghiu/inflation}},\ \Eprint
  {https://arxiv.org/abs/2211.04483} {arXiv:2211.04483} \BibitemShut {NoStop}%
\bibitem [{\citenamefont {Navascués}\ and\ \citenamefont
  {Wolfe}(2020)}]{navascues2017}%
  \BibitemOpen
  \bibfield  {author} {\bibinfo {author} {\bibfnamefont {M.}~\bibnamefont
  {Navascués}}\ and\ \bibinfo {author} {\bibfnamefont {E.}~\bibnamefont
  {Wolfe}},\ }\bibfield  {title} {\bibinfo {title} {{T}he inflation technique
  completely solves the causal compatibility problem},\ }\href
  {https://doi.org/10.1515/jci-2018-0008} {\bibfield  {journal} {\bibinfo
  {journal} {J. Causal Inference}\ }\textbf {\bibinfo {volume} {8}},\ \bibinfo
  {pages} {70} (\bibinfo {year} {2020})},\ \Eprint
  {https://arxiv.org/abs/1707.06476} {arXiv:1707.06476} \BibitemShut {NoStop}%
\bibitem [{\citenamefont {Ligthart}\ \emph {et~al.}(2023)\citenamefont
  {Ligthart}, \citenamefont {Gachechiladze},\ and\ \citenamefont
  {Gross}}]{ligthart2023}%
  \BibitemOpen
  \bibfield  {author} {\bibinfo {author} {\bibfnamefont {L.~T.}\ \bibnamefont
  {Ligthart}}, \bibinfo {author} {\bibfnamefont {M.}~\bibnamefont
  {Gachechiladze}},\ and\ \bibinfo {author} {\bibfnamefont {D.}~\bibnamefont
  {Gross}},\ }\bibfield  {title} {\bibinfo {title} {A convergent inflation
  hierarchy for quantum causal structures},\ }\href
  {https://doi.org/10.1007/s00220-023-04697-7} {\bibfield  {journal} {\bibinfo
  {journal} {Commun. Math. Phys.}\ }\textbf {\bibinfo {volume} {401}},\
  \bibinfo {pages} {2673} (\bibinfo {year} {2023})},\ \Eprint
  {https://arxiv.org/abs/2110.14659} {arXiv:2110.14659} \BibitemShut {NoStop}%
\bibitem [{\citenamefont {Ligthart}\ and\ \citenamefont
  {Gross}(2023)}]{ligthart2023b}%
  \BibitemOpen
  \bibfield  {author} {\bibinfo {author} {\bibfnamefont {L.~T.}\ \bibnamefont
  {Ligthart}}\ and\ \bibinfo {author} {\bibfnamefont {D.}~\bibnamefont
  {Gross}},\ }\bibfield  {title} {\bibinfo {title} {The inflation hierarchy and
  the polarization hierarchy are complete for the quantum bilocal scenario},\
  }\href {https://doi.org/10.1063/5.0143792} {\bibfield  {journal} {\bibinfo
  {journal} {J. Math. Phys.}\ }\textbf {\bibinfo {volume} {64}},\ \bibinfo
  {pages} {072201} (\bibinfo {year} {2023})},\ \Eprint
  {https://arxiv.org/abs/2212.11299} {arXiv:2212.11299} \BibitemShut {NoStop}%
\bibitem [{\citenamefont {Girardin}\ and\ \citenamefont
  {Gisin}(2023)}]{girardin2023}%
  \BibitemOpen
  \bibfield  {author} {\bibinfo {author} {\bibfnamefont {A.}~\bibnamefont
  {Girardin}}\ and\ \bibinfo {author} {\bibfnamefont {N.}~\bibnamefont
  {Gisin}},\ }\bibfield  {title} {\bibinfo {title} {Violation of the {F}inner
  inequality in the four-output triangle network},\ }\href
  {https://doi.org/10.1103/PhysRevA.108.042213} {\bibfield  {journal} {\bibinfo
   {journal} {Phys. Rev. A}\ }\textbf {\bibinfo {volume} {108}},\ \bibinfo
  {pages} {042213} (\bibinfo {year} {2023})},\ \Eprint
  {https://arxiv.org/abs/2306.05922} {arXiv:2306.05922} \BibitemShut {NoStop}%
\bibitem [{\citenamefont {Terhal}(2000)}]{Terhal2000}%
  \BibitemOpen
  \bibfield  {author} {\bibinfo {author} {\bibfnamefont {B.~M.}\ \bibnamefont
  {Terhal}},\ }\bibfield  {title} {\bibinfo {title} {Bell inequalities and the
  separability criterion},\ }\href
  {https://doi.org/https://doi.org/10.1016/S0375-9601(00)00401-1} {\bibfield
  {journal} {\bibinfo  {journal} {Phys. Lett. A}\ }\textbf {\bibinfo {volume}
  {271}},\ \bibinfo {pages} {319} (\bibinfo {year} {2000})},\ \Eprint
  {https://arxiv.org/abs/quant-ph/9911057} {arXiv:quant-ph/9911057}
  \BibitemShut {NoStop}%
\bibitem [{\citenamefont {Lewenstein}\ \emph {et~al.}(2001)\citenamefont
  {Lewenstein}, \citenamefont {Kraus}, \citenamefont {Horodecki},\ and\
  \citenamefont {Cirac}}]{Lewenstein2001}%
  \BibitemOpen
  \bibfield  {author} {\bibinfo {author} {\bibfnamefont {M.}~\bibnamefont
  {Lewenstein}}, \bibinfo {author} {\bibfnamefont {B.}~\bibnamefont {Kraus}},
  \bibinfo {author} {\bibfnamefont {P.}~\bibnamefont {Horodecki}},\ and\
  \bibinfo {author} {\bibfnamefont {J.~I.}\ \bibnamefont {Cirac}},\ }\bibfield
  {title} {\bibinfo {title} {Characterization of separable states and
  entanglement witnesses},\ }\href {https://doi.org/10.1103/PhysRevA.63.044304}
  {\bibfield  {journal} {\bibinfo  {journal} {Phys. Rev. A}\ }\textbf {\bibinfo
  {volume} {63}},\ \bibinfo {pages} {044304} (\bibinfo {year} {2001})},\
  \Eprint {https://arxiv.org/abs/quant-ph/0005112} {arXiv:quant-ph/0005112}
  \BibitemShut {NoStop}%
\bibitem [{\citenamefont {Chruściński}\ and\ \citenamefont
  {Sarbicki}(2014)}]{Chruscinski2014}%
  \BibitemOpen
  \bibfield  {author} {\bibinfo {author} {\bibfnamefont {D.}~\bibnamefont
  {Chruściński}}\ and\ \bibinfo {author} {\bibfnamefont {G.}~\bibnamefont
  {Sarbicki}},\ }\bibfield  {title} {\bibinfo {title} {Entanglement witnesses:
  construction, analysis and classification},\ }\href
  {https://doi.org/10.1088/1751-8113/47/48/483001} {\bibfield  {journal}
  {\bibinfo  {journal} {J. Phys. A: Math. Theor.}\ }\textbf {\bibinfo {volume}
  {47}},\ \bibinfo {pages} {483001} (\bibinfo {year} {2014})},\ \Eprint
  {https://arxiv.org/abs/1402.2413} {arXiv:1402.2413} \BibitemShut {NoStop}%
\bibitem [{\citenamefont {Brunner}\ \emph {et~al.}(2014)\citenamefont
  {Brunner}, \citenamefont {Cavalcanti}, \citenamefont {Pironio}, \citenamefont
  {Scarani},\ and\ \citenamefont {Wehner}}]{Brunner2014}%
  \BibitemOpen
  \bibfield  {author} {\bibinfo {author} {\bibfnamefont {N.}~\bibnamefont
  {Brunner}}, \bibinfo {author} {\bibfnamefont {D.}~\bibnamefont {Cavalcanti}},
  \bibinfo {author} {\bibfnamefont {S.}~\bibnamefont {Pironio}}, \bibinfo
  {author} {\bibfnamefont {V.}~\bibnamefont {Scarani}},\ and\ \bibinfo {author}
  {\bibfnamefont {S.}~\bibnamefont {Wehner}},\ }\bibfield  {title} {\bibinfo
  {title} {Bell nonlocality},\ }\href
  {https://doi.org/10.1103/RevModPhys.86.419} {\bibfield  {journal} {\bibinfo
  {journal} {Rev. Mod. Phys.}\ }\textbf {\bibinfo {volume} {86}},\ \bibinfo
  {pages} {419} (\bibinfo {year} {2014})},\ \Eprint
  {https://arxiv.org/abs/1303.2849} {arXiv:1303.2849} \BibitemShut {NoStop}%
\bibitem [{\citenamefont {Designolle}\ \emph {et~al.}(2024)\citenamefont
  {Designolle}, \citenamefont {V\'ertesi},\ and\ \citenamefont
  {Pokutta}}]{Designolle2024}%
  \BibitemOpen
  \bibfield  {author} {\bibinfo {author} {\bibfnamefont {S.}~\bibnamefont
  {Designolle}}, \bibinfo {author} {\bibfnamefont {T.}~\bibnamefont
  {V\'ertesi}},\ and\ \bibinfo {author} {\bibfnamefont {S.}~\bibnamefont
  {Pokutta}},\ }\bibfield  {title} {\bibinfo {title} {{Symmetric multipartite
  Bell inequalities via Frank-Wolfe algorithms}},\ }\href
  {https://doi.org/10.1103/PhysRevA.109.022205} {\bibfield  {journal} {\bibinfo
   {journal} {Phys. Rev. A}\ }\textbf {\bibinfo {volume} {109}},\ \bibinfo
  {pages} {022205} (\bibinfo {year} {2024})},\ \Eprint
  {https://arxiv.org/abs/2310.20677} {arXiv:2310.20677} \BibitemShut {NoStop}%
\bibitem [{\citenamefont {Ulibarrena}\ \emph {et~al.}(2024)\citenamefont
  {Ulibarrena}, \citenamefont {Webb}, \citenamefont {Pickston}, \citenamefont
  {Ho}, \citenamefont {Fedrizzi},\ and\ \citenamefont
  {Pozas-Kerstjens}}]{Ulibarrena2024}%
  \BibitemOpen
  \bibfield  {author} {\bibinfo {author} {\bibfnamefont {A.}~\bibnamefont
  {Ulibarrena}}, \bibinfo {author} {\bibfnamefont {J.~W.}\ \bibnamefont
  {Webb}}, \bibinfo {author} {\bibfnamefont {A.}~\bibnamefont {Pickston}},
  \bibinfo {author} {\bibfnamefont {J.}~\bibnamefont {Ho}}, \bibinfo {author}
  {\bibfnamefont {A.}~\bibnamefont {Fedrizzi}},\ and\ \bibinfo {author}
  {\bibfnamefont {A.}~\bibnamefont {Pozas-Kerstjens}},\ }\bibfield  {title}
  {\bibinfo {title} {Guarantees on the structure of experimental quantum
  networks},\ }\href {https://doi.org/10.1038/s41534-024-00911-z} {\bibfield
  {journal} {\bibinfo  {journal} {npj Quantum Inf.}\ }\textbf {\bibinfo
  {volume} {10}},\ \bibinfo {pages} {117} (\bibinfo {year} {2024})},\ \Eprint
  {https://arxiv.org/abs/2403.02376} {arXiv:2403.02376} \BibitemShut {NoStop}%
\bibitem [{\citenamefont {Pozas-Kerstjens}(2019)}]{AlexThesis}%
  \BibitemOpen
  \bibfield  {author} {\bibinfo {author} {\bibfnamefont {A.}~\bibnamefont
  {Pozas-Kerstjens}},\ }\emph {\bibinfo {title} {Quantum information outside
  quantum information}},\ \href {http://hdl.handle.net/10803/667696} {Ph.D.
  thesis},\ \bibinfo  {school} {Universitat Polit\`ecnica de Catalunya}
  (\bibinfo {year} {2019})\BibitemShut {NoStop}%
\bibitem [{\citenamefont {Plávala}\ \emph {et~al.}(2025)\citenamefont
  {Plávala}, \citenamefont {Ligthart},\ and\ \citenamefont
  {Gross}}]{plavala2025}%
  \BibitemOpen
  \bibfield  {author} {\bibinfo {author} {\bibfnamefont {M.}~\bibnamefont
  {Plávala}}, \bibinfo {author} {\bibfnamefont {L.~T.}\ \bibnamefont
  {Ligthart}},\ and\ \bibinfo {author} {\bibfnamefont {D.}~\bibnamefont
  {Gross}},\ }\bibfield  {title} {\bibinfo {title} {The polarization hierarchy
  for polynomial optimization over convex bodies, with applications to
  nonnegative matrix rank},\ }\href
  {https://doi.org/https://doi.org/10.1016/j.laa.2025.05.019} {\bibfield
  {journal} {\bibinfo  {journal} {Linear Algebra Appl.}\ }\textbf {\bibinfo
  {volume} {723}},\ \bibinfo {pages} {15} (\bibinfo {year} {2025})},\ \Eprint
  {https://arxiv.org/abs/2406.09506} {arXiv:2406.09506} \BibitemShut {NoStop}%
\bibitem [{\citenamefont {Frisch}(1995)}]{frisch1995turbulence}%
  \BibitemOpen
  \bibfield  {author} {\bibinfo {author} {\bibfnamefont {U.}~\bibnamefont
  {Frisch}},\ }\href@noop {} {\emph {\bibinfo {title} {Turbulence: the legacy
  of {A. N. Kolmogorov}}}}\ (\bibinfo  {publisher} {Cambridge University
  Press},\ \bibinfo {year} {1995})\BibitemShut {NoStop}%
\bibitem [{\citenamefont {Shavit}\ \emph {et~al.}(2022)\citenamefont {Shavit},
  \citenamefont {Vladimirova},\ and\ \citenamefont {Falkovich}}]{shavit2022}%
  \BibitemOpen
  \bibfield  {author} {\bibinfo {author} {\bibfnamefont {M.}~\bibnamefont
  {Shavit}}, \bibinfo {author} {\bibfnamefont {N.}~\bibnamefont
  {Vladimirova}},\ and\ \bibinfo {author} {\bibfnamefont {G.}~\bibnamefont
  {Falkovich}},\ }\bibfield  {title} {\bibinfo {title} {Emerging scale
  invariance in a model of turbulence of vortices and waves},\ }\href
  {https://doi.org/10.1098/rsta.2021.0080} {\bibfield  {journal} {\bibinfo
  {journal} {Phil. Trans. R. Soc. A}\ }\textbf {\bibinfo {volume} {380}},\
  \bibinfo {pages} {20210080} (\bibinfo {year} {2022})}\BibitemShut {NoStop}%
\end{thebibliography}%

\appendix
\onecolumngrid
\setcounter{figure}{0}
\renewcommand{\thefigure}{\thesection\arabic{figure}}

\section{The example distribution and its triangle-local model}
\label{app:E1c}
The distribution that we use throughout the work is initially described in the Supplemental Material of Ref.~\cite{gisin2020}.
It corresponds to the distribution with $E_2=-1/3$ that has the highest value of $E_1$ while being producible when the latent nodes represent sources of physical systems described by classical mechanics.
This value, which we call $E_1^c$ in the main text, is
\begin{equation*}
    \begin{aligned}
        E_1^c=&\,\frac{1}{2} 
        -\frac{1}{6\sqrt{\frac{3}{3\ 2^{5/3}\sqrt[3]{3\sqrt{41} + 25} + \sqrt[3]{21600 - 2592\sqrt{41}} - 21}}} \\
        &+ \frac{1}{2}\sqrt{\frac{34}{\sqrt{3\left(3\ 2^{5/3}\sqrt[3]{3\sqrt{41} + 25} + \sqrt[3]{21600 - 2592\sqrt{41}} - 21 \right)}}-\frac{1}{9} 2^{5/3}\sqrt[3]{3\sqrt{41} + 25} - \frac{1}{27}\sqrt[3]{21600 - 2592\sqrt{41}} - \frac{14}{9}}\\
        \approx&\,0.1753.
   \end{aligned}
\end{equation*}

When $E_2=-1/3$, the positivity of all probabilities requires that $E_3=-3E_1$.
Thus, the distribution that we will consider has $E_3=E_3^c\approx-0.5260$.

This distribution can be produced in the triangle scenario when the latent nodes distribute information encoded in systems that follow the laws of classical physics.
A mechanism for producing it is given in Supplementary Note 2 of Ref.~\cite{gisin2020}, which we reproduce here for completeness.
In the notation given in Ref.~\cite{gisin2020}, the local model is
\begin{equation*}
	P = \left[
	\begin{array}{c|ccc}
		& 1 & 2 & 3\\
		\hline
		\alpha & x & 1-x & 0\\
		\beta & y & (1-y)/2 & (1-y)/2\\
		\gamma & 1-x & x & 0
	\end{array} \right], \quad
	f_a =
	\begin{bmatrix}
		1 & 0 & 1\\
		0 & 0 & 0\\
		1 & 1 & 1\\
	\end{bmatrix}, \quad
	f_b =
	\begin{bmatrix}
		1 & 1 & 0 \\
		0 & 1 & 0\\
		0 & 0 & 0
	\end{bmatrix},\quad
	f_c =
	\begin{bmatrix}
		0 & 1 & 0 \\
		1 & 1 & 0\\
		0 & 0 & 0
	\end{bmatrix},
\end{equation*}
where $x$ is the root between $0$ and $1$ for $3x^4-9x^3+9x^2-5x+1=0$, i.e.,
\begin{align*}
    x=&\,\frac{3}{4}-\frac {1} {2}\sqrt{\frac{1}{2} + \frac{2}{3}\sqrt[3]{\frac{2}{\sqrt{41}+3}} - \frac13\sqrt[3]{\frac{1}{2}\left(\sqrt{41} + 3\right)} + \frac{13}{2\sqrt{3\left(3-8\sqrt[3]{\frac{2}{\sqrt{41}+3}} + 2^{5/3}\sqrt[3]{\sqrt{41}+3}\right)}}} \\
    &+\frac{1}{4\sqrt{\frac{3}{3-8\sqrt[3]{\frac{2}{\sqrt{41} + 3}} + 2^{5/3}\sqrt[3]{\sqrt{41} + 3}}}}\\
    \approx&\,0.3579
\end{align*}
and $y=\frac{1}{3(2x^2-2x+1)}\approx 0.6168$.

\section{Witnesses of incompatibility from variable elimination}\label{app:elimination}
We build a certificate of incompatibility with the seven-sided inflation in Fig.~\ref{fig:inflation} by performing variable elimination on the positivity constraints $\{p_\text{inf}(a_0,\dots,a_6)\geq 0\}_{a_0,\dots,a_6}$ using positive coefficients only.
In this case, this is a set of $2^7=128$ constraints.
Requiring that the probabilities are compatible with the inflation imposes the constraints \eqref{eq:symm}-\eqref{eq:iden2}.
Equation \eqref{eq:symm} reduces the number of positivity constraints to only 20, while Eqs.~\eqref{eq:iden1} and \eqref{eq:iden2} relate the variables and allow us to perform the variable elimination.

In order to work with these constraints, it turns out to be simpler to work in terms of correlators, which are given by the discrete Fourier transform of the probabilities, i.e.,
\begin{equation}
    p(a_0,\dots,a_6)=\frac{1}{2^7}\left(1+a_0 F_{A_0} + a_1 F_{A_1} + \dots + a_0 a_1 F_{A_0A_1} + a_0 a_2 F_{A_0A_2} + \dots+a_0a_1a_2a_3a_4a_5a_6F_{A_0A_1A_2A_3A_4A_5A_6}\right),
\end{equation}
where $F_j\in[-1,1]$ are the correlators.
In this notation, the symmetry constraint \eqref{eq:symm} is reflected in the fact that many correlators are equal, i.e., $F_{A_0}=F_{A_1}=\dots=F_1$, $F_{A_0A_1}=F_{A_1A_2}=\dots=F_2$, and so on.
Importantly, $F_{A_0A_2}=F_{1,1}$, which needs not be equal to $F_2$.
The constraints \eqref{eq:iden1} and \eqref{eq:iden2} are reflected in two ways.
First, correlators that involve distant parties factorize, i.e., $F_{1,1}=F_1^2$ and so on.
Second, since the single- and two-body marginals of the inflation coincide with those in the original network, we have that $F_1=E_1$ and $F_2=E_2$.
In summary, we have that the distributions that can be generated in the seven-sided inflation of Fig.~\ref{fig:inflation} can be parametrized in terms of just five variables, $\{F_3,F_4,F_5,F_6,F_7\}$, via
\begin{equation}
    \begin{aligned}
        p(a_0,&\dots,a_6)=\frac{1}{2^7} \left[1\right.\\
        + &\,(a_0 + a_1 + a_2 + a_3 + a_4 + a_5 + a_6) E_1 \\
        + &\,(a_0 a_1 + a_1 a_2 + a_2 a_3 + a_3 a_4 + a_4 a_5 + a_5 a_6 + a_6 a_0) E_2 \\
        + &\,\left(a_0 (a_2 + a_3 + a_4 + a_5) + a_1 (a_3 + a_4 + a_5 + a_6) + a_2 (a_4 + a_5 + a_6) + a_3 (a_5 + a_6) + a_4 a_6\right) E_1^2 \\
        + &\,(a_0 a_1 a_2 + a_1 a_2 a_3 + a_2 a_3 a_4 + a_3 a_4 a_5 + a_4 a_5 a_6 + a_5 a_6 a_0 + a_6 a_0 a_1) F_3 \\
        + &\,\left(a_0 a_1 (a_3 + a_4 + a_5) + a_1 a_2 (a_4 + a_5 + a_6) + a_2 a_3 (a_5 + a_6 + a_0) \right.\\
        &+\left. a_3 a_4 (a_6 + a_0 + a_1) + a_4 a_5 (a_0 + a_1 + a_2) + a_5 a_6 (a_1 + a_2 + a_3) + a_6 a_0 (a_2 + a_3 + a_4)\right) E_1 E_2 \\
        + &\,\left(a_0 (a_2 (a_4 + a_5) + a_3 a_5) + a_1 (a_3 (a_5 + a_6) + a_4 a_6) + a_2 a_4 a_6\right) E_1^3 \\
        + &\,(a_0 a_1 a_2 a_3 + a_1 a_2 a_3 a_4 + a_2 a_3 a_4 a_5 + a_3 a_4 a_5 a_6 + a_4 a_5 a_6 a_0 + a_5 a_6 a_0 a_1 + a_6 a_0 a_1 a_2) F_4 \\
        + &\,\left(a_0 a_1 a_2 (a_4 + a_5) + a_1 a_2 a_3 (a_5 + a_6) + a_2 a_3 a_4 (a_6 + a_0) + a_3 a_4 a_5 (a_0 + a_1) \right. \\
        &+\left. a_4 a_5 a_6 (a_1 + a_2) + a_5 a_6 a_0 (a_2 + a_3) + a_6 a_0 a_1 (a_3 + a_4)\right) F_3 E_1 \\
        + &\,\left(a_0 a_1 (a_3 a_4 + a_4 a_5) + a_1 a_2 (a_4 a_5 + a_5 a_6) + a_2 a_3 (a_5 a_6 + a_6 a_0) + a_3 a_4 a_6 a_0\right) E_2^2 \\
        + &\,(a_0 a_1 a_3 a_5 + a_1 a_2 a_4 a_6 + a_2 a_3 a_5 a_0 + a_3 a_4 a_6 a_1 + a_4 a_5 a_0 a_2 + a_5 a_6 a_1 a_3 + a_6 a_0 a_2 a_4) E_2 E_1^2 \\
        + &\,(a_0 a_1 a_2 a_3 a_4 + a_1 a_2 a_3 a_4 a_5 + a_2 a_3 a_4 a_5 a_6 + a_3 a_4 a_5 a_6 a_0 + a_4 a_5 a_6 a_0 a_1 + a_5 a_6 a_0 a_1 a_2 + a_6 a_0 a_1 a_2 a_3) F_5 \\
        + &\,(a_0 a_1 a_2 a_3 a_5 + a_1 a_2 a_3 a_4 a_6 + a_2 a_3 a_4 a_5 a_0 + a_3 a_4 a_5 a_6 a_1 + a_4 a_5 a_6 a_0 a_2 + a_5 a_6 a_0 a_1 a_3 + a_6 a_0 a_1 a_2 a_4) F_4 E_1 \\
        + &\,(a_0 a_1 a_2 a_4 a_5 + a_1 a_2 a_3 a_5 a_6 + a_2 a_3 a_4 a_6 a_0 + a_3 a_4 a_5 a_0 a_1 + a_4 a_5 a_6 a_1 a_2 + a_5 a_6 a_0 a_2 a_3 + a_6 a_0 a_1 a_3 a_4) F_3 E_2 \\
        + &\,(a_0 a_1 a_2 a_3 a_4 a_5\!+\!a_1 a_2 a_3 a_4 a_5 a_6\!+\!a_2 a_3 a_4 a_5 a_6 a_0\!+\!a_3 a_4 a_5 a_6 a_0 a_1\!+\!a_4 a_5 a_6 a_0 a_1 a_2\!+\!a_5 a_6 a_0 a_1 a_2 a_3\!+\!a_6 a_0 a_1 a_2 a_3 a_4) F_6 \\
        + &\left. a_0 a_1 a_2 a_3 a_4 a_5 a_6 F_7\right].
    \end{aligned}
    \label{eq:7probs}
\end{equation}

\begin{figure}
    \centering
    \includegraphics[width=0.6\linewidth]{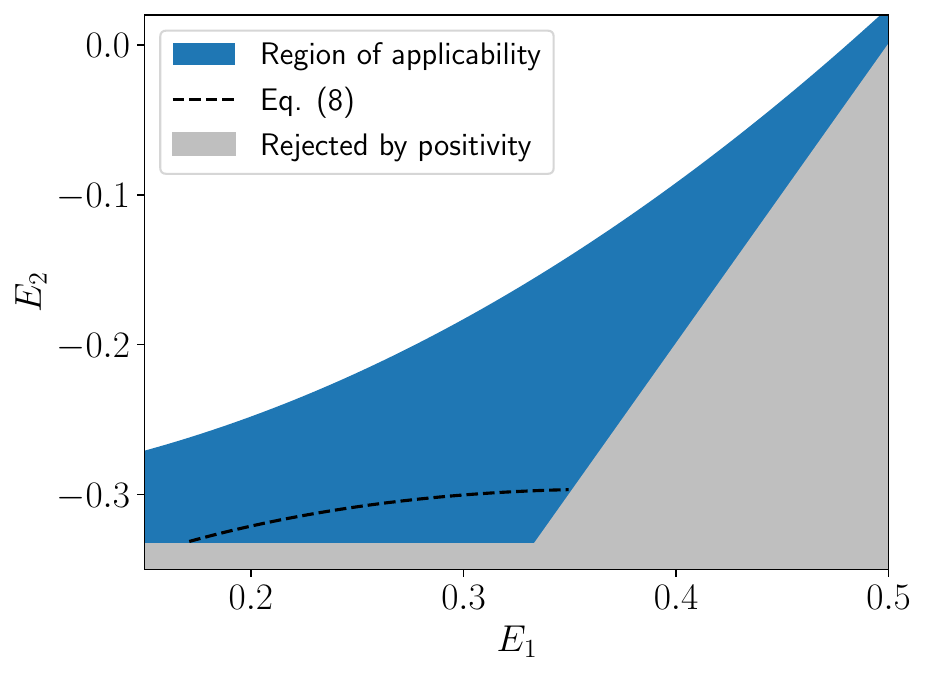}
    \caption{\textbf{Region of applicability of Eq.~\eqref{eq:7cert}.} Depicted is the region in $E_1-E_2$ space where the conditions $6E_1+5E_2+1\geq0$, $13E_1-1\geq0$, and $13E_1^2-10E_2-3\geq0$ are satisfied.
    These are necessary conditions for Eq.~\eqref{eq:7cert} to be a certificate of incompatibility with symmetric realizations.
    Thus, all the distributions below the dashed line are certified not to admit a realization in terms of symmetric strategies.}
    \label{fig:7cert}
\end{figure}

Thus, we search for linear combinations of the constraints $\{p_\text{inf}(a_0,\dots,a_6)\geq 0\}$ with positive coefficients that eliminate the variables $\{F_3,F_4,F_5,F_6,F_7\}$.
In doing so, we allow for the coefficients to be not just non-negative real numbers, but also functions of $E_1$ and $E_2$ that are non-negative in the neighborhood of the distribution of interest.
Using Gauss-Jordan elimination, we arrive at the following combination (for simplicity, we use the notation $p_{a_0\dots a_6}=p_\text{inf}(a_0,\dots,a_6)$):
\begin{equation}
    \begin{aligned}
        \mathcal{C}=\frac14(E_1 + 1) (6 E_1 + 5 E_2 + 1) &\bigg[28 p_{++++++-} + 70 p_{+++----} + 21 p_{+-+----} + 21 p_{-------} \\
       & + \left.\frac{1}{2} (13 E_1 - 1) \left(5 p_{+++++++} + 7 p_{++++++-} + 7 p_{+------} + 5 p_{-------}\right)\right] \\
        & \hspace{-3.8cm}+ \frac{7}{8}(13 E_1^2 - 10 E_2 - 3) \left[(E_1 + 1) \left(4 p_{++++++-} + 10 p_{+++----} + 3 p_{+-+----} + 3 p_{-------}\right)\right. \\
        & + \left.(13 E_1 - 1) (p_{+++----} + 2 p_{++-----} + p_{+------})\right].
    \end{aligned}
\end{equation}
Whenever the quantities $6E_1+5E_2+1$, $13E_1-1$, and $13E_1^2-10E_2-3$ are non-negative (which is a large part of the region we are interested in; see Fig.~\ref{fig:7cert}), the above is a linear combination of probabilities with positive coefficients and thus positive for any $p_\text{inf}$ that can be produced in the inflation in Fig.~\ref{fig:inflation}.
When replacing the probabilities by Eq.~\eqref{eq:7probs}, we arrive at the witness of incompatibility in Eq.~\eqref{eq:7cert}.

\section{The witness of infeasibility from linear programming}
\label{app:cert}
In this appendix we show the witness of infeasibility that limits the dark green region in Fig.~\ref{fig:results}.
Its violation by a symmetric distribution, i.e., characterized by the parameters $E_1$, $E_2$ and $E_3\in[-1,1]$ via Eq.~\eqref{eq:prob}, is a proof that the distribution does not admit any realization in the triangle scenario where the three states distributed are the same and the parties all perform the same measurement.
The witness is obtained via Farkas' lemma on the linear program associated with the 15th level of the inflation hierarchy described in the main text and takes the form
\begin{equation}
    \begin{aligned}
    0&.1843 E_1^9 - 1.8290 E_1^8 - 4.6758 E_1^7 E_2 - 39.5446 E_1^7 - 0.8972 E_1^6 E_2^2 - 74.0838 E_1^6 E_2 - 116.2647 E_1^6 \\
    &+ 10.5142 E_1^5 E_2^2 - 248.3040 E_1^5 E_2 - 79.1817 E_1^5 - 28.6167 E_1^4 E_2^3 - 238.3961 E_1^4 E_2^2 - 532.6388 E_1^4 E_2 \\
    &- 153.0877 E_1^4 - 3.3372 E_1^3 E_2^4 - 43.4770 E_1^3 E_2^3 - 451.7175 E_1^3 E_2^2 - 308.6530 E_1^3 E_2 + 32.9430 E_1^3 \\
    &+ 18.5457 E_1^2 E_2^4 - 73.4824 E_1^2 E_2^3 - 491.2832 E_1^2 E_2^2 - 450.5083 E_1^2 E_2 - 135.2657 E_1^2 + 1.3657 E_1 E_2^5 \\
    &- 35.9763 E_1 E_2^4 - 225.1462 E_1 E_2^3 - 310.1144 E_1 E_2^2 - 75.8387 E_1 E_2 + 13.3812 E_1 - 0.2856 E_2^6 \\
    &- 4.6748 E_2^5 - 34.0812 E_2^4 - 75.4711 E_2^3 - 123.2455 E_2^2 - 79.4846 E_2 - 16.5823 \leq 0.
    \end{aligned}
\end{equation}
The code that obtains this witness can be found in the computational appendix \cite{compapp}.

\end{document}